\documentclass[pra,aps,showpacs,showkeys,groupedaddress,superscriptaddress,twocolumn]{revtex4-1}

\usepackage[utf8x]{inputenc}
\usepackage{color}
\usepackage{bbm} 
\usepackage{amsfonts,amsmath,amssymb,stmaryrd}
\usepackage{graphicx}
\usepackage{CJK}
\usepackage{subfigure} 
\usepackage{hyperref}
\usepackage{epsfig}
\usepackage{mathrsfs}
\usepackage{verbatim}
\usepackage{centernot}
\usepackage{stackengine}

\newcommand{\ket}[1]{|#1\rangle}

\usepackage{array}

\usepackage{bm}	

\begin{document}
\title{Robust double Bragg diffraction via detuning control}

\author{Rui Li}
\author{V. J. Mart\'inez-Lahuerta}
\affiliation{Leibniz University Hanover, Institute of Quantum Optics, Hannover, Germany}
\author{S. Seckmeyer}
\affiliation{Leibniz University Hanover, Institute of Quantum Optics, Hannover, Germany}
\author{Klemens Hammerer}
\affiliation{Leibniz University Hanover, Institute of Theoretical Physics, Hannover, Germany}
\author{Naceur Gaaloul}
\affiliation{Leibniz University Hanover, Institute of Quantum Optics, Hannover, Germany}

\begin{abstract}
We present a theoretical model and numerical optimization of double Bragg diffraction, a widely used technique in atom interferometry. We derive an effective two-level-system Hamiltonian based on the Magnus expansion in the so-called ``quasi-Bragg regime", where most Bragg-pulse atom interferometers operate. Furthermore, we extend the theory to a five-level description to account for Doppler detuning. Using these derived effective Hamiltonians, we investigate the impacts of AC-Stark shift and polarization errors on the double Bragg beam-splitter, along with their mitigations through detuning control. Notably, we design a linear detuning sweep that demonstrates robust efficiency exceeding 99.5\% against polarization errors up to 8.5\%. Moreover, we develop an artificial intelligence-aided optimal detuning control protocol, showcasing enhanced robustness against both polarization errors and Doppler effects. This protocol achieves an average efficiency of 99.92\% for samples with a finite momentum width of $0.05\,\hbar k_L$ within an extended polarization error range of up to 10\%.
\end{abstract}
   

\date{\today}
\maketitle 
\section{Introduction}
Atom interferometry (AI) has emerged as a powerful tool for precision measurements~\cite{Chu-Kasevich-1991, Chu-Kasevich-1992, Fixler-science-2007, Lamporesi-PRL-2008, Parker-Science-2018, Morel-Nature-2020}, inertial navigation~\cite{Geiger-2011,Cheiney-PRApp-2018} and gravitational wave detection~\cite{Graham-PRL-2013, Canuel-SciRep-2018, Canuel-2020, Zhan-JMP-2020, MAGIS-100-2021}. Achieving high-precision measurements in AI relies on either large momentum transfer (LMT)~\cite{Gebbe-NatCom-2021, Wilkason-PRL-2022, Rodzinka-2024} or extended interrogation times~\cite{Müntinga-2013, Xu-Science-2019, Asenbaum-2020, Panda-Nat-Phys-2024}. LMT between atoms and light can be realized through various methods such as standing light waves~\cite{Rasel-PRL-1995}, single or double Raman diffraction~\cite{Chu-Kasevich-1991, Lévèque-2009} or Bragg diffraction~\cite{Torii-PRA-2000, Gebbe-NatCom-2021, Müntinga-2013}, often followed by Bloch oscillations~\cite{Cladé-2009, Müller-2009}. Double Bragg diffraction (DBD), which utilizes two counter-propagating optical lattices to diffract atoms, is an LMT technique employed in modern AI experiments. This technique effectively enhances sensitivity by doubling the scale factor relative to single Bragg diffraction (SBD). Most importantly, DBD possesses an intrinsic symmetry crucial for improving measurement accuracy by suppression of phase noise and systematic uncertainties~\cite{Ahlers-PRL-2016}, making it particularly suited for microgravity and space experiments~\cite{Altschul-2015, Barrett-2016, Zoest-2010, Becker-Nature-2018, Gaaloul-NatComm-2022, Elliott-Nature-2023, CARIOQA-2023}. 

Previous theoretical studies of double Bragg diffraction have focused on a perturbative approach aimed at adiabatically eliminating higher-order diffractions and yielding a reduced set of differential equations describing the lowest lying momentum states~\cite{Giese-PRA-2013, Giese2015}. 
This approach has been applied to describe recent DBD experiments and has shown good agreement~\cite{Ahlers-PRL-2016}. However, the method of averaging, as used in prior literature, requires a transformation from the momentum-state picture to a dressed-state picture. This complicates the connection to the initial conditions and measured signals, which both have simple forms in the bare momentum states.

In this work, we develop an alternative approach for describing DBD based on a Magnus expansion~\cite{Magnus-1954, Blanes-2009} of the DBD Hamiltonian. This allows us to account for time-dependent Rabi frequencies and lattice imperfections due to unavoidable polarization errors. Based on this, we develop several control strategies to achieve high-efficiency beam splitters (BS) despite these imperfections. We complement our analytical approach for describing DBD with numerical optimizations based on optimal control theory (OCT)~\cite{OCTInQuantumMechanicalSystems1,OCTAlgorithms}, which has been effective in designing robust AI pulses, primarily in the context of SBD~\cite{KovachyAI,QCTRLAI,OptimalControlFountainsKasevich,OptimalControlToulouse}.

To gain a deeper understanding of the physics of DBD within the quasi-Bragg regime, considering time-dependent Rabi frequencies, polarization errors, and Doppler effects, we develop a theoretical framework in Sec.~\ref{II} by exploiting the inherent  symmetry of DBD and the Magnus expansion, and further extend this framework to include Doppler detuning. In Sec.~\ref{III}, we explore mitigation methods for AC-Stark shift and polarization errors using constant, linear, or OCT-optimized detuning functions. Finally, in Sec.~\ref{IV}, we address losses and asymmetry induced by Doppler effects and propose various mitigation strategies based on detuning control. 
 
\section{Theoretical framework}\label{II}
\subsection{Symmetry of the double Bragg Hamiltonian}
\begin{figure}[t]
  \centering
  \includegraphics[width=1.0\columnwidth]{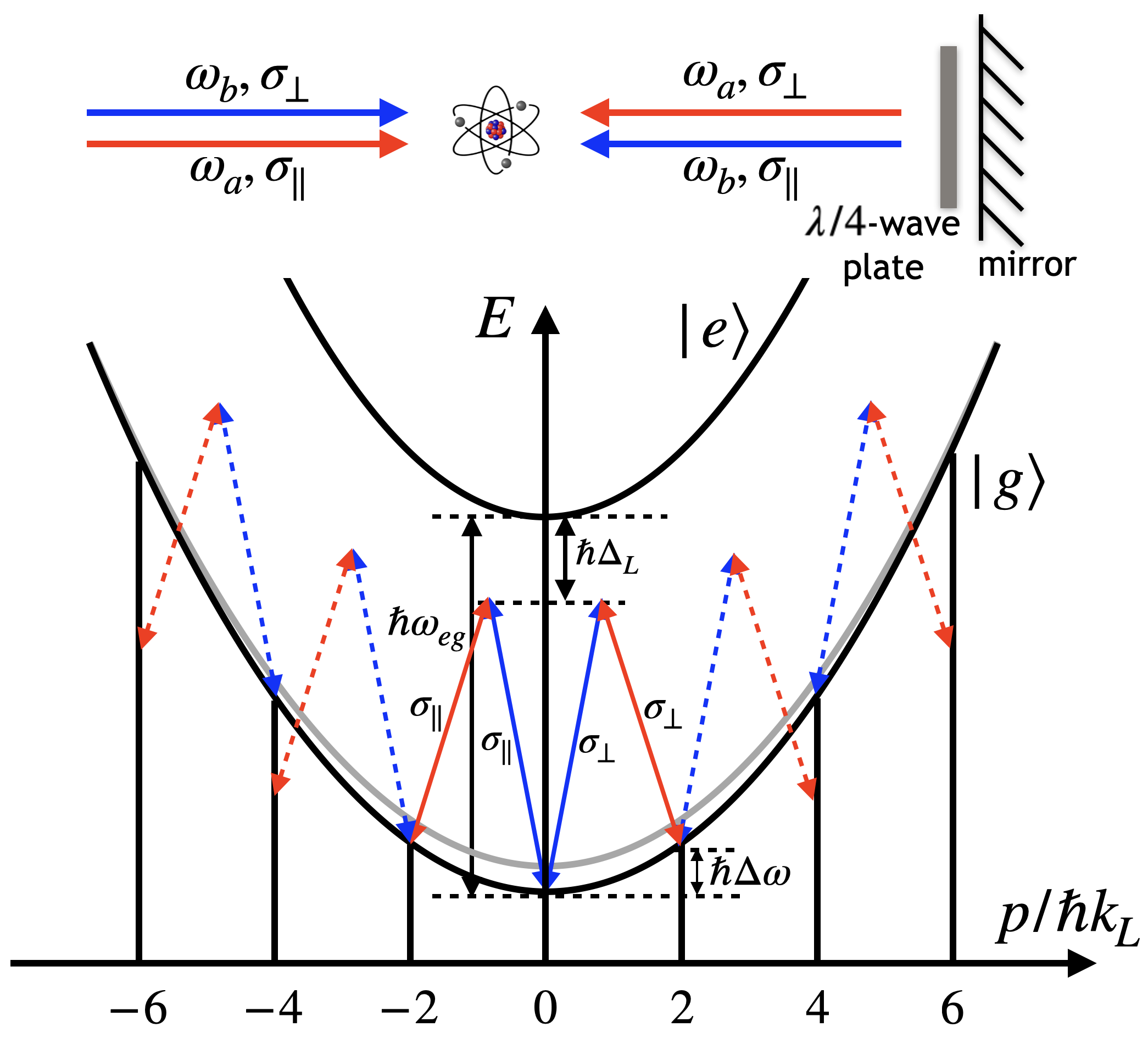}
    \caption{Realization of DBD via a retro-reflection setup (upper part) and energy diagram (lower part) showing the first-order resonance condition. Here, $\Delta \omega  =  \omega_b- \omega_a= 4\,\omega_{rec} +\Delta$ is the frequency difference of the two lasers with detuning $\Delta$ from the traditional resonance condition. $\sigma_\perp$ and $\sigma_\parallel$ denote the polarizations of the two input beams, which should be orthogonal to each other ideally. $\Delta_L$ is the laser detuning to the $|g\rangle \rightarrow |e\rangle$ transition. The ground state energy dispersion illustrated in solid-black (grey) has (has not) taken into account corrections due to the AC-Stark shift.}
    \label{fig: DBD}
 \end{figure}

We consider atoms diffracted by two pairs of counter-propagating laser beams with different frequencies, $\omega_a$ and $\omega_b$. Each pair is associated with a unique polarization, either $\sigma_\perp$ or $\sigma_\parallel$, and ideally, their polarizations are orthogonal to each other (see Fig.~\ref{fig: DBD}). 
The one-dimensional (1D) real-space Hamiltonian describing the above double Bragg diffraction can be written as:
\begin{align}
     H(t)=&\frac{\hat p^2}{2m} +2\hbar \Omega(t) \Big\{\epsilon_{pol}\big[\cos(2k_L x) + \cos\big(\Delta\omega(t)t \big)\big]\nonumber\\&+\cos^2(k_L x + \frac{\Delta \omega(t)}{2}t) + \cos^2(k_L x - \frac{\Delta \omega(t)}{2}t)\Big\} \nonumber\\
    =&\frac{\hat p^2}{2m} +2\hbar \Omega(t) \cos(2 k_L x) \Big\{\cos\big[\Delta\omega(t) t\big]+\epsilon_{pol}\Big\}\nonumber\\
    &+\text{Const}_x\mathbf{\hat 1},\label{eq:1}
\end{align}
where $\epsilon_{pol}=|\sigma_\perp\cdot\sigma_{\parallel}|$ is the polarization error due to imperfect beam polarizations, and $\Omega(t)=\Omega_a(t) \Omega_b(t)/(2\Delta_L)$ is a time-dependent two-photon Rabi frequency, with $\Omega_a(t)$ and $\Omega_b(t)$ being single-photon Rabi frequencies. $\Delta \omega (t) = \omega_b-\omega_a = 4\,\omega_{rec} +\Delta(t)$ is the time-dependent frequency difference between the two input beams~\cite{footnote_on_omega}, with  $\omega_{rec}=\hbar k_L^2/(2m)$ being the single-photon recoil frequency. Fig.~\ref{fig: DBD} illustrates the energy diagram of DBD including transitions and one possible realization of the Hamiltonian~\eqref{eq:1} via retro-reflection of two laser beams with frequencies $\omega_a$ and $\omega_b$ and linear polarizations $\sigma_\perp$ and $\sigma_\parallel$, respectively~\cite{Ahlers-PRL-2016, Abend-PRL-2016}. We assume that a double pass of the $\lambda/4$-wave plate, as shown in Fig.~\ref{fig: DBD}, switches the two beam polarizations by adjusting the $\lambda/4$-wave plate's fast axis. Ideally, the two counter-propagating optical lattices are orthogonal in their polarizations, each driving an SBD. 

Evidently, the double Bragg Hamiltonian~(\ref{eq:1}) possesses a \textit{parity symmetry}, given by $\hat{\mathbf{P}}:x\rightarrow - x$ in real space, and $[\hat{\mathbf{P}},  H(t)]=0$ at all times. Therefore, the full Hilbert space splits into a direct sum of even and odd subspaces:
$\mathcal{H}=\mathcal{H}_{even}\oplus \mathcal{H}_{odd}$. For an ideal momentum eigenstate with zero momentum (at rest with respect to the mirror in Fig.~\ref{fig: DBD} which defines the frame of reference for the DBD Hamiltonian~\eqref{eq:1}) as the initial state, i.e., $|\psi(t=0)\rangle =|0\hbar k_L\rangle$, the wave function will only evolve in the subspace $\mathcal{H}_{even}$ spanned by the even momentum basis:
\begin{align}
|n\rangle = \left\{\begin{array}{ll}
|0\hbar k_L\rangle, & n=0  \\
\frac{1}{\sqrt{2}}(|2n\hbar k_L\rangle + |-2n\hbar k_L\rangle), & n>0 
\end{array}
\right. 
\end{align}
Thus, the full Hamiltonian $H(t)$ in the even momentum basis can be expressed as 
\begin{align}
H(t) =& \sum_{n=0}^\infty 4n^2\hbar \omega_{rec}|n\rangle \langle n|+\hbar \Omega(t) C(t, \epsilon_{pol})\times\Big\{\nonumber\\&\sqrt{2}\big(|0\rangle \langle 1| + |1\rangle \langle 0|\big)+\sum_{n=1}^\infty\big(|n\rangle \langle n+1| + |n+1\rangle \langle n|\big)\Big\}\nonumber\\
=& H_0 +H_1(t)\label{eq:2},
\end{align}
where $C(t, \epsilon_{pol})=\cos\big[\Delta \omega (t)t\big]+\epsilon_{pol}$, and we have divided the full Hamiltonian into the time-independent part $H_0$ and the time-dependent part $H_1(t)$ and neglected the constant part in Hamiltonian~(\ref{eq:1}) by shifting the zero-point energy. 
In the interaction picture with respect to the diagonal Hamiltonian $H_0= \sum_{n=0}^\infty 4n^2\hbar \omega_{rec}|n\rangle \langle n|$, one obtains 

\begin{align}
\bar H(t)=& e^{i\frac{H_0}{\hbar}t}H_1(t) e^{-i\frac{H_0}{\hbar}t}\nonumber\\
=&\hbar\Omega(t)C(t, \epsilon_{pol})\times\Big\{\sum_{n=1}^\infty\big(e^{i 4 (2n+1)\omega_{rec} t}|\overline{n+1}\rangle \langle \overline{n}| \nonumber\\&+ e^{-i 4 (2n+1)\omega_{rec} t}|\overline{n}\rangle \langle \overline{n+1}|\big)\nonumber\\&+\sqrt{2}(e^{-i 4\omega_{rec}t}|\overline{0}\rangle\langle\overline{1}| + e^{i 4\omega_{rec}t}|\overline{1}\rangle\langle\overline{0}|)\Big\}\label{eq:H_int},
\end{align}
when expressed in the transformed even momentum basis $|\bar n\rangle = e^{i\frac{H_0}{\hbar}t} | n\rangle $. We will not distinguish between $| n\rangle$ and $|\bar n\rangle$ in the following text as they only differ by a pure phase which has no effect on the probabilities.

\subsection{Magnus expansion and effective two-level Hamiltonian}
The interaction Hamiltonian $\bar H(t)$ given by Eq.~(\ref{eq:H_int}) serves as the foundation for deriving the effective description for the DBD without Doppler detuning in this section.  
For a system evolving under a generic time-dependent Hamiltonian $ \bar{H}(t)$, the unitary time evolution operator $U(t,\,0)$ satisfies the Schrödinger equation 
\begin{equation}
 i\hbar \frac{d}{dt}U(t,\,0)=\bar{H}(t)\,U(t,\,0) ,   
\end{equation}
whose solution can be written as a matrix exponential of the \textit{Magnus expansion} \cite{Magnus-1954, Blanes-2009}:
\begin{align}
    U(t,\,0)&=\exp\Big[\sum_{i=1}^{+\infty}G_i(t)\Big]. \label{Eq:Magnus_1}
\end{align}
The first two terms in the Magnus expansion are given by  
\begin{align}
    G_1(t)&=-\frac{i}{\hbar}\int_0^t \bar H(t_1) \,d t_1,\\
    G_2(t)&=\Big(-\frac{i}{\hbar}\Big)^2\frac{1}{2!}\int_0^t\!\!\! dt_1 \int_0^{t_1} \!\!\!dt_2\big[\bar H(t_1),\bar H(t_2)\big].
\end{align}
Our procedure is to truncate the Magnus expansion~(\ref{Eq:Magnus_1}) at the second order and define an effective Hamiltonian by
\begin{align}
    H_{\text{eff}}(t)&:=i\hbar \frac{d}{dt}\big[G_1(t) + G_2(t)\big],
\end{align}
from which we can gain physical insights into the dynamics. We also use it to solve the Schrödinger equation in real-time, i.e., constructing the unitary time evolution operator with a Dyson series~\cite{Dyson-1949}:
\begin{align}
U(t,\,0) = \mathcal{T}\exp\Big[-\frac{i}{\hbar} \int_{0}^{t} H_{\text{eff}}(t) \,dt\Big].\label{Eq:Magnus_2}
\end{align} 
Moreover, we adiabatically eliminate all higher-order transitions as well as fast oscillations in the quasi-Bragg regime $\Omega(t), |\Delta| \ll 8\,\omega_{rec}$. This process yields the following effective two-level-system (TLS) Hamiltonian:
\begin{widetext}
\begin{align}
    H_{eff}=& \hbar \left(\begin{matrix}
\frac{\Omega^2}{\omega_{rec}} \big(\frac{1}{4}\epsilon_{pol} - \frac{1}{2}\epsilon_{pol}^2\big) & \frac{\sqrt{2}}{2}\Omega\big\{e^{i \Delta t}+e^{-i (\Delta +8\omega_{rec}) t}+2\,\epsilon_{pol}e^{-i 4\omega_{rec}t}\big\} \\
\frac{\sqrt{2}}{2}\Omega\big\{e^{-i \Delta  t}+e^{i (\Delta +8\omega_{rec}) t}+2\,\epsilon_{pol}e^{i 4\omega_{rec} t}\big\}  & \frac{\Omega^2}{\omega_{rec}}\big(-\frac{3}{64}-\frac{1}{4}\epsilon_{pol}+\frac{5}{12}\epsilon^2_{pol}\big)\\
\end{matrix} \right)\label{eq:H_eff},
\end{align}
where the basis used is $\{|\bar 0 \rangle,\,|\bar 1 \rangle\}$ introduced in Eq.~\eqref{eq:H_int}. A time-dependent detuning $\Delta=\Delta \omega(t)-4\,\omega_{rec}$ is introduced and $\Omega=\Omega(t)$ is the time-dependent Rabi frequency. In the second-order Magnus expansion, $\Delta$ is assumed to be a small constant and the limit $\Delta\rightarrow 0$ is taken. 
\end{widetext}
The $(1,1)$ and $(2,2)$-matrix elements arise from the second-order Magnus expansion, while the rest comes from the first-order Magnus expansion. In the TLS Hamiltonian~(\ref{eq:H_eff}), the $(1,1)$ and $(2,2)$-matrix elements represent light-induced second-order energy shifts of the state $|0\rangle$ and $|1\rangle$, respectively. We term this energy shift tensor as the time-dependent \textit{AC-Stark shift} ($\Delta_{AC}$) of the first-order double Bragg diffraction. 

It should be noted that the above procedure involves a certain level of double-counting regarding the effect of the AC-Stark shift. We have also investigated other options and found the differences to be marginal. We emphasize that this simplified two-level picture of the effective Hamiltonian is intended to provide insight into the essence of the dynamics (such as the shift of the DBD resonance or temporal control of the detuning) rather than to serve as a replacement for the exact Hamiltonian~\eqref{eq:1}.

\subsection{Five-level description with Doppler detuning}\label{Section_5-LS}
To understand the DBD of atomic wave packets with a finite momentum spread, we assume an input Gaussian wave function in momentum space given by
\begin{align}
    \psi(p)=\left(2\pi\sigma^2_p\right)^{-1/4}\exp\left(-\frac{(p-p_0)^2}{4 \sigma_p^2}\right),
\end{align}
where $p_0$ is the center-of-mass (COM) momentum of the atoms with respect to the mirror in Fig.~\ref{fig: DBD}, and $\sigma_p$ is the momentum width which can be associated with a pseudo 1D temperature defined by $m k_B T \equiv \sigma_p^2$ along the DBD beam axis, where $k_B$ is Boltzmann's constant. The initial state expressed in the plane-wave basis is given by
\begin{align}
    |\psi(t=0)\rangle=\int \psi(p)\,|p\rangle \,d p,
\end{align}
where $|p\rangle$ denotes a momentum eigenstate with normalization condition $\langle p'|p\rangle =\delta(p-p')$. Since it is a superposition of different momentum eigenstates, one only needs to know how a particular momentum eigenstate $|p\rangle$ will evolve under the full DBD Hamiltonian~(\ref{eq:1}), and then superpose the resulting wave functions to obtain the final outcome state, i.e.,
\begin{align}
    |\psi(t=t)\rangle=\hat U(t,0) \,|\psi(t=0)\rangle=\int \psi(p)\,\hat U(t,0)|p\rangle \,d p. 
\end{align}

The DBD Hamiltonian~\eqref{eq:1} is notably invariant under discrete translation given by $\mathbf{\hat T}_n : x \rightarrow x+n\pi/k_L $, which allows us to divide the full momentum space into Brillouin zones, i.e., $\mathcal{P}=\cup_{n=-\infty}^{+\infty}[-\hbar k_L+2n\hbar k_L,\,\hbar k_L+2n\hbar k_L)$. We consider an ideal atomic wave packet with a definite initial momentum from the first Brillouin zone $p\in [-\hbar k_L,\, \hbar k_L)$. With this choice, our initial state reads $|\psi(t=0)\rangle=|p\rangle$. For DBD up to $\pm 4\hbar k_L$ momentum transfer, there are five relevant basis vectors:
\begin{align}
    |1\rangle&=|p\rangle,\nonumber\\|2 \text{ or } 3\rangle&=\frac{|p+2\hbar k_L\rangle \pm |p-2\hbar k_L\rangle}{\sqrt{2}},\nonumber\\ 
    |4\text{ or }5\rangle&=\frac{|p+4\hbar k_L\rangle \pm |p-4\hbar k_L\rangle}{\sqrt{2}}\label{Eq: five bases}.
\end{align}
It is worth noting that none of the five basis vectors belongs to either even or odd subspace for $p\neq0$. Hence, we will only refer to them as symmetric and anti-symmetric states for the rest of the paper. In the above truncated basis, the full DBD Hamiltonian~(\ref{eq:1}) (neglecting the constant part), with polarization errors $\epsilon_{pol}$ and Doppler detuning due to non-zero initial momentum $p$, can be expressed as the following $5\times 5$ matrix:
\begin{widetext}
\begin{align}
   H_{Doppler}^{pol}/\hbar = \left(\begin{matrix}
\frac{p^2}{2m\hbar} &  \sqrt{2}\Omega(t)C(t, \epsilon_{pol})  &0&0 & 0\\
\sqrt{2}\Omega(t)C(t, \epsilon_{pol})  &\frac{p^2}{2m\hbar} + 4\omega_{rec}& 4\frac{p}{\hbar k_L}\omega_{rec}&\Omega(t)C(t, \epsilon_{pol})  & 0\\
 0 &4\frac{p}{\hbar k_L}\omega_{rec}  &\frac{p^2}{2m\hbar} + 4 \omega_{rec}&0&\Omega(t)C(t, \epsilon_{pol})\\
 0&\Omega(t)C(t, \epsilon_{pol})&0&\frac{p^2}{2m\hbar} + 16\omega_{rec}&8\frac{p}{\hbar k_L}\omega_{rec}\\
 0& 0&\Omega(t)C(t, \epsilon_{pol})&8\frac{p}{\hbar k_L}\omega_{rec}&\frac{p^2}{2m\hbar} + 16 \omega_{rec}
\end{matrix} \right)\label{eq: H_full},
\end{align}
where the basis is given by Eq. (\ref{Eq: five bases}) and $C(t, \epsilon_{pol})=\cos\big[(4\,\omega_{rec}+\Delta(t))t\big]+\epsilon_{pol}$.
\end{widetext}
 We note that the Doppler shift $4\,p\,\omega_{rec} /(\hbar k_L)=2\,k_Lp/m=2\,k_L v$, given by the matrix elements $(2,3)$ and $(3,2)$, results in a direct coupling between the symmetric state $|2\rangle$ and the anti-symmetric state $|3\rangle$. In the bare momentum basis, given by $|p\pm2\hbar k_L\rangle=(|2\rangle\pm|3\rangle)/\sqrt{2}$, it amounts to an opposite phase evolution of each momentum state. Therefore, for $p \neq 0$, we choose to characterize the efficiency of double Bragg diffraction in terms of the populations in the bare momentum states. The AC-Stark shift of state $|2\rangle$ and $|3\rangle$ due to the virtual population of $|4\rangle$ and $|5\rangle$ is automatically taken care of by solving the 5-level Hamiltonian~(\ref{eq: H_full}) in real-time. For the full Hamiltonian with Doppler detuning, we refer the readers to Appendix~\ref{A}.
\section{Mitigations of AC-Stark Shift and Polarization Errors} \label{III}
In this section, we apply our theory to the two most widely used pulse shapes in DBD experiments: box and Gaussian pulses. Although the traditional double-Bragg resonance condition for the ideal case of vanishing initial momentum in the literature reads $\Delta\omega = (2j\hbar k_L)^2/(2 j m\hbar) = 4j\,\omega_{rec}$, where $j$ stands for $j$th-order DBD~\cite{Giese-PRA-2013, Giese2015}, we will show that the traditional resonance condition breaks down due to the time-dependent AC-Stark shift to both initial and target states. Consequently, achieving a perfect population inversion from the initial state $|0\rangle$ to the target state $|1\rangle$ (a perfect BS) by simply setting $\Delta \omega=4\,\omega_{rec}$, even without any polarization error in the system, becomes unattainable.  However, we propose that one can compensate for this AC-Stark shift as well as polarization errors and still achieve resonant DBD using either a constant or a general time-dependent detuning function $\Delta(t) = \Delta \omega (t)-4\,\omega_{rec}$. All results obtained in this section and the following sections assume a plane-wave initial state (except for the exact numerical solutions, which use a finite but very narrow Gaussian momentum distribution with a width of $\sigma_p=0.01 \hbar k_L$, centered at $p_0=p$ to approximate the plane-wave initial state $|p\rangle$) unless otherwise stated.
\subsection{AC-Stark shifted DBD resonance condition}
\begin{figure}[t]
    \centering
    \subfigure[]{\includegraphics[width=0.49\columnwidth]{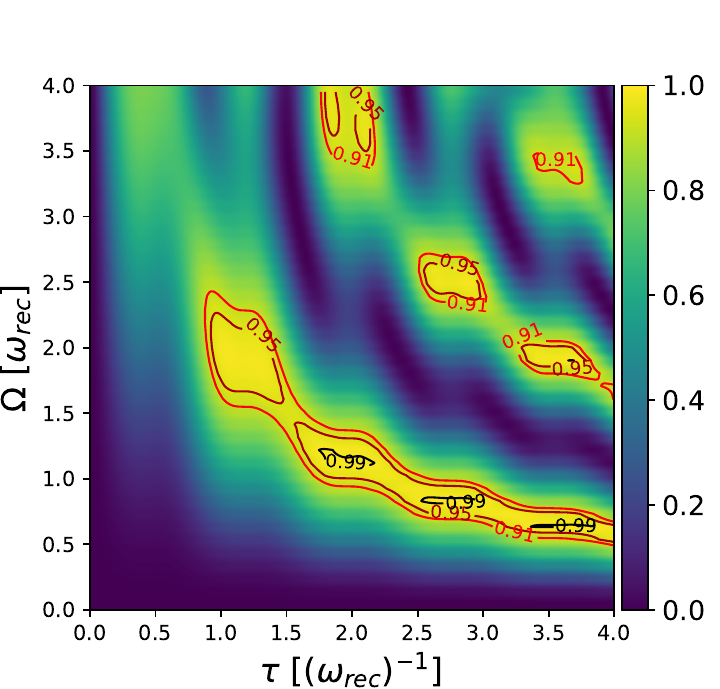}} 
    \subfigure[]{\includegraphics[width=0.49\columnwidth]{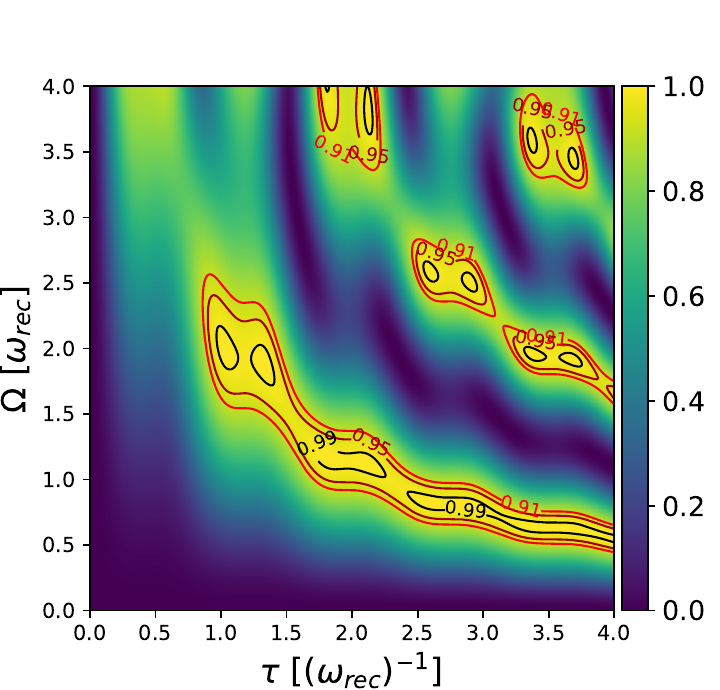}} 
    \caption{DBD efficiency for a box-pulse Rabi frequency as a function of box-pulse duration $\tau$ and peak Rabi frequency $\Omega$ with $\epsilon_{pol}=0$ and $\Delta=0$. (a) Exact dynamics described by the full Hamiltonian (\ref{eq:1}) calculated with UATIS with an initial momentum width $\sigma_p=0.01\,\hbar k_L$ and mean momentum $p_0=0$. (b) Dynamics described by the effective TLS Hamiltonian~(\ref{eq:H_eff}) calculated by numerically solving the truncated Schrödinger equation. The black and red contour lines show regions with a DBD efficiency exceeding $91\%$.}
    \label{Fig: 2D_scan_box}
\end{figure}

To demonstrate the modified double-Bragg resonance condition due to the AC-Stark shift, we apply our effective two-level theory to the simple case of a box pulse with no polarization error ($\epsilon_{pol}=0$). We define the DBD efficiency as the target state population $P(|1\rangle)$ after a box pulse given by 
\begin{align}
    \Omega(t)=\Omega\,\big[\Theta(t)-\Theta(t-\tau)\big],
\end{align} 
where $\Theta(t)$ is the Heaviside step function. We show the $(\Omega,\, \tau)$-parameter scans of the DBD efficiency in Fig.~\ref{Fig: 2D_scan_box} and highlight the regions with a DBD efficiency exceeding $91\%$ by contour plots. Fig.~\ref{Fig: 2D_scan_box}a shows the results of the exact dynamics predicted by the full Hamiltonian~(\ref{eq:1}) calculated by a position-space numerical solver of the Schrödinger equation (UATIS \cite{UATIS}) based on the second-order Suzuki-Trotter decomposition~\cite{Suzuki-1990} with a momentum truncation up to $\pm10.9\,\hbar k_L$, which we will refer to as the ``exact numerical solution", or simply ``exact" in the figure legends. Fig.~\ref{Fig: 2D_scan_box}b shows the dynamics predicted by our effective two-level Hamiltonian~($\ref{eq:H_eff}$). Both are calculated for $\Delta=0$ and $\epsilon_{pol}=0$. 
We find that the 2D patterns of the DBD efficiency given by the exact numerical solution and the effective two-level theory are in good qualitative agreement. 
\begin{figure}[h]
  \centering
  \includegraphics[width=\columnwidth]{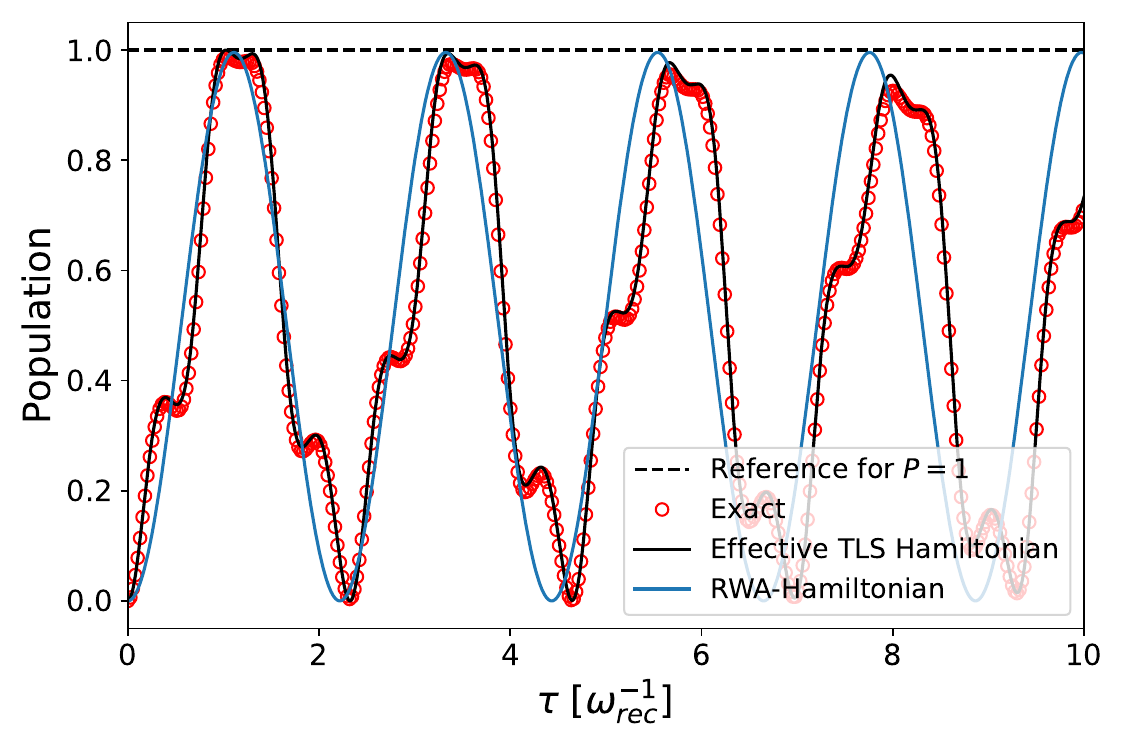}
    \caption{Time evolution of excited state population $P(|1\rangle)$ with initial state $|\psi(t=0)\rangle =|0\rangle$ as a function of box-pulse duration $\tau $ for a constant Rabi frequency $\Omega=2\,\omega_{rec}$, governed by the exact real-space Hamiltonian (\ref{eq:1}) (red circles), the effective TLS Hamiltonian (\ref{eq:H_eff}) (black curve) and the RWA-Hamiltonian (\ref{eq: RWA}) (blue curve) for $\epsilon_{pol}=0$ and $\Delta=0$. }
    \label{Fig:compare_three}
\end{figure}

For a quantitative comparison, we plot the time evolution of $P(|1\rangle)$ along the horizontal cut $\Omega = 2\,\omega_{rec}$ up to $\tau =10\, \omega_{rec}^{-1}$ in Fig.~\ref{Fig:compare_three}. Additionally, we also plot the curve predicted by the pulse-area condition for DBD as given in the literature~\cite{Giese-PRA-2013, Giese2015}, which we will derive as a limiting case of our effective TLS Hamiltonian~($\ref{eq:H_eff}$). To recover the pulse-area condition,  we first transform the TLS Hamiltonian~($\ref{eq:H_eff}$) into an interaction picture with respect to $H_0=\text{diag}(0,\,\Delta)$ for a constant detuning $\Delta$. Then, we apply a rotating wave approximation (RWA) to eliminate every oscillating term on the off-diagonals, resulting in a time-independent Hamiltonian:
\begin{align}
   \bar H_{\text{RWA}}= \hbar \left(\begin{matrix}
0 & \frac{\sqrt{2}}{2}\Omega \\
\frac{\sqrt{2}}{2}\Omega  & \delta_{diff}\\
\end{matrix} \right),  \label{eq: RWA}
\end{align}
which we refer to as the RWA-Hamiltonian. In Eq.~\eqref{eq: RWA}, we have defined the total differential light shift of DBD due to frequency detuning $\Delta$ and AC-Stark shift $\Delta_{AC}$ as
\begin{align}
\delta_{diff}= -\Delta -\frac{3}{64} \frac{\Omega^2}{\omega_{rec}}\label{eq:diff_shift},
\end{align}
which simply vanishes for DBD operating at the traditional DBD resonance. The general solution of this TLS Hamiltonian with the initial condition $c_n(t=0)=\delta_{n,0}$ and time-independent $\Omega$ is given by \textit{Rabi's formula}:
\begin{align}
    P_{|0\rangle\rightarrow |1\rangle}(t)=\frac{2\Omega^2}{2\Omega^2 + \delta_{diff}^2}\sin^2\Big(\frac{\sqrt{2\Omega^2 +\delta_{diff}^2}}{2}t\Big)\label{eq:Rabi's formular}.
\end{align}
From Rabi's formula, it can be deduced that the pulse-area condition is given by $\sqrt{2\Omega^2 +\delta_{diff}^2} t=\pi$ and a perfect double Bragg BS (a full population inversion) is only reachable for $\delta_{diff} = 0$, where the pulse-area condition reduces to $\sqrt{2}\Omega t=\pi$. 

The dynamics predicted by Eq.~($\ref{eq:Rabi's formular}$) for $\delta_{diff}=0$ (traditional resonance condition) are illustrated by the blue curve in Fig.~\ref{Fig:compare_three}, where the time $t$ is replaced by the box-pulse duration $\tau$. We compare this against the exact dynamics (red circles) and those governed by the effective TLS Hamiltonian~(\ref{eq:H_eff}) (black curve) in the same figure. It is evident that the RWA-Hamiltonian, as given by Eq.~(\ref{eq: RWA}), and the consequent pulse-area condition fail to capture some important features of the time evolution of $P(|1\rangle)$, such as higher-order oscillations due to the counter-rotating terms, as evidenced by the clear discrepancy  between the exact dynamics (red circles) and the results of the RWA-Hamiltonian (blue curve) in Fig.~\ref{Fig:compare_three}. On the other hand, the results of our effective TLS Hamiltonian (black curve) show excellent agreement with the exact dynamics (red circles) with the largest deviation noted to be within $3\%$ for $\tau \in [0, 10\, \omega_{rec}^{-1}]$. Hence, we conclude that for a comprehensive physical description of box-pulse DBD in the quasi-Bragg regime, the time-dependent effective Hamiltonian~(\ref{eq:H_eff}) should be adopted instead of the RWA-Hamiltonian~($\ref{eq: RWA}$). 

 \subsection{Polarization errors and mitigations}\label{sec_pol_err_mitigation}
As shown in the $(2,2)$-matrix element of the TLS Hamiltonian (\ref{eq:H_eff}), polarization errors also contribute to the AC-Stark shift of the target state. Therefore, we aim to simultaneously mitigate both effects via a suitable control of the detuning. While our study of DBD with polarization errors will primarily focus on the Gaussian pulses, similar results can also be obtained in the case of box pulses. The time-dependent Rabi frequency of Gaussian-pulse DBD is given by
\begin{align}
    \Omega(t)= \Omega_R\, e^{-\frac{(t-t_0)^2}{2\tau^2}}\label{eq:gaussian},
\end{align}
with the peak Rabi frequency $\Omega_R$, temporal center of the Gaussian $t_0$ and the pulse width $\tau$. We always set $t_0=0$ by adjusting the laser phase unless specified otherwise. The pulse-area condition for Gaussian pulses, without AC-Stark shift and detuning ($\delta_{diff}=0$), can be deduced from Rabi's formula (\ref{eq:Rabi's formular}) as follows:
\begin{align}
     P_{|0\rangle\rightarrow |1\rangle}(\tau)=\sin^2\Big(\frac{\sqrt{2}}{2}\int_{-\infty}^{+\infty}\Omega(t) dt\Big)=\sin^2\Big(\sqrt{\pi}\Omega_R\tau \Big)\label{eq:pulse-area},
\end{align}
where the condition for a full population inversion is given by $\Omega_R\tau=\sqrt{\pi}/2$. 
\begin{figure}[t]
  \centering
  \includegraphics[width=1.0\columnwidth]{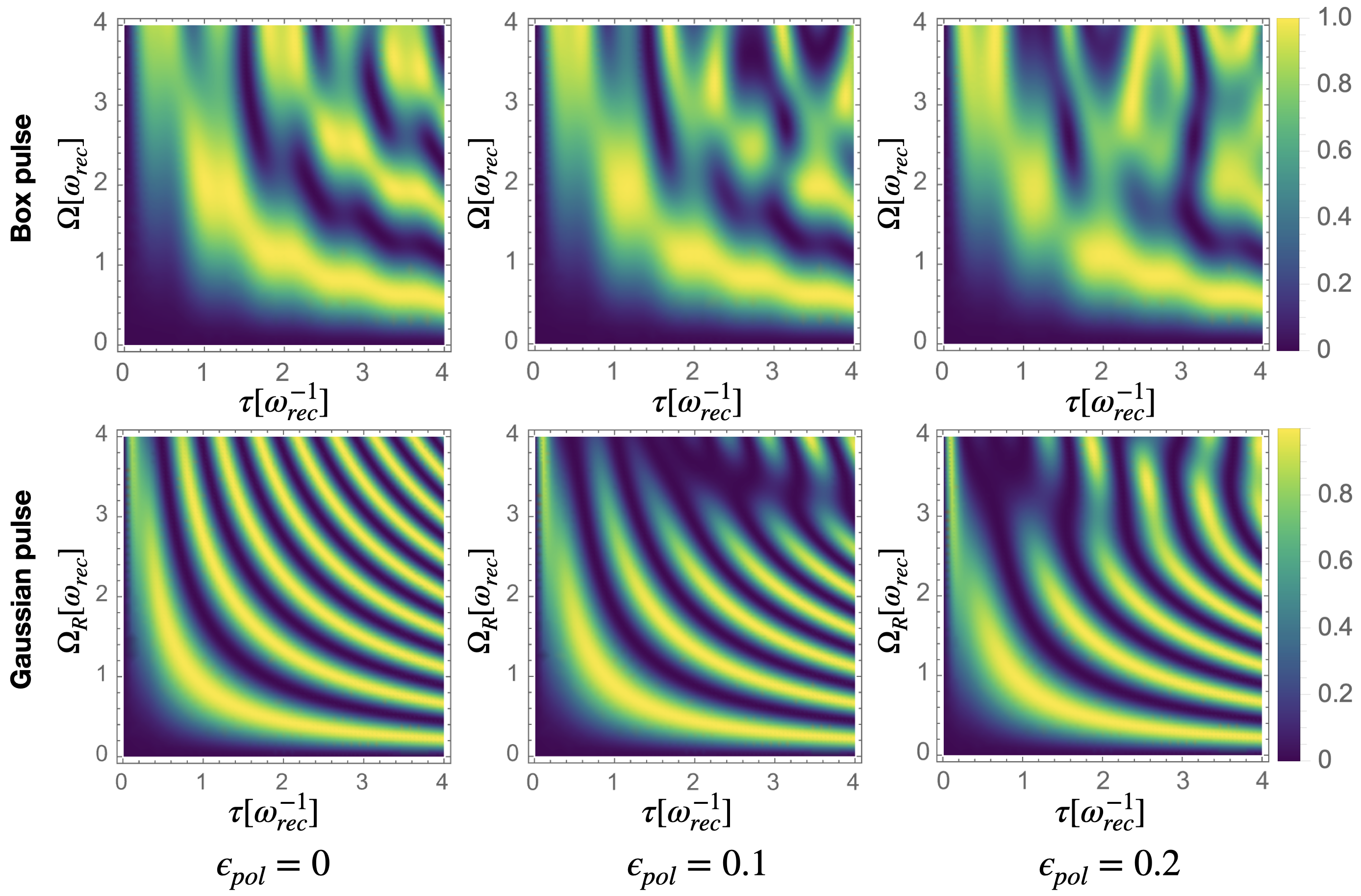}
    \caption{DBD efficiency for a box-pulse (upper row) or Gaussian-pulse (lower row) with various polarization errors $\epsilon_{pol}=(0,\,0.1,\,0.2)$ in the left, middle and right column predicted by the effective TLS Hamiltonian~(\ref{eq:H_eff}) as a function of the pulse width and the peak Rabi frequency ($\tau,\,\Omega\text{ or }\Omega_R$) with $\Delta=0$.}
    \label{fig:2D_pol_scan}
 \end{figure}
 
We numerically solve the effective time-dependent TLS Hamiltonian~(\ref{eq:H_eff}) and obtain a 2D parameter scan of Gaussian-pulse DBD efficiency as a function of $(\Omega_R,\tau)$ for various polarization errors in the lower row of Fig.~\ref{fig:2D_pol_scan}. A similar 2D scan is also calculated for box-pulse DBD, shown in the upper row of Fig.~\ref{fig:2D_pol_scan}, solely for qualitative comparison. A detailed comparison to the exact numerical solution can be found in Appendix \ref{box_pol}. We find it better to work with Gaussian-pulse DBD, as it has less distortion in the first three Rabi cycles under polarization errors and is closer to an ideal two-level system. 
 
\begin{figure}[t]
  \centering
  \subfigure[]{\includegraphics[width=1.0\columnwidth]{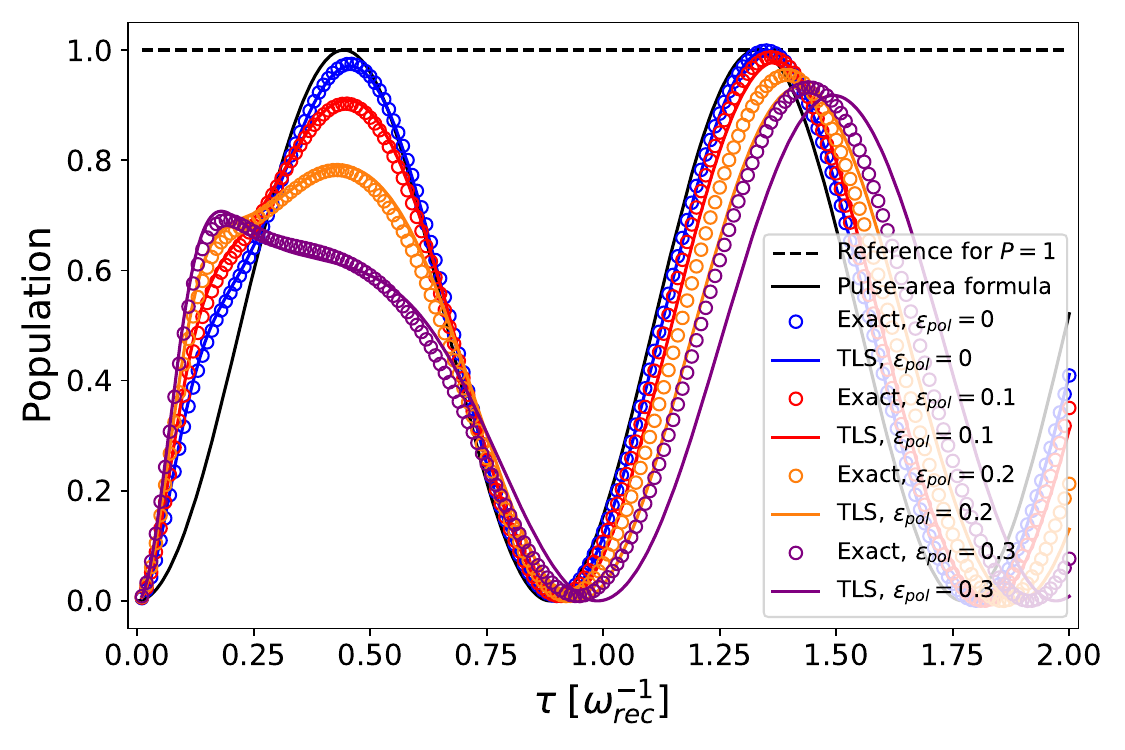}} 
  \subfigure[]{\includegraphics[width=1.0\columnwidth]{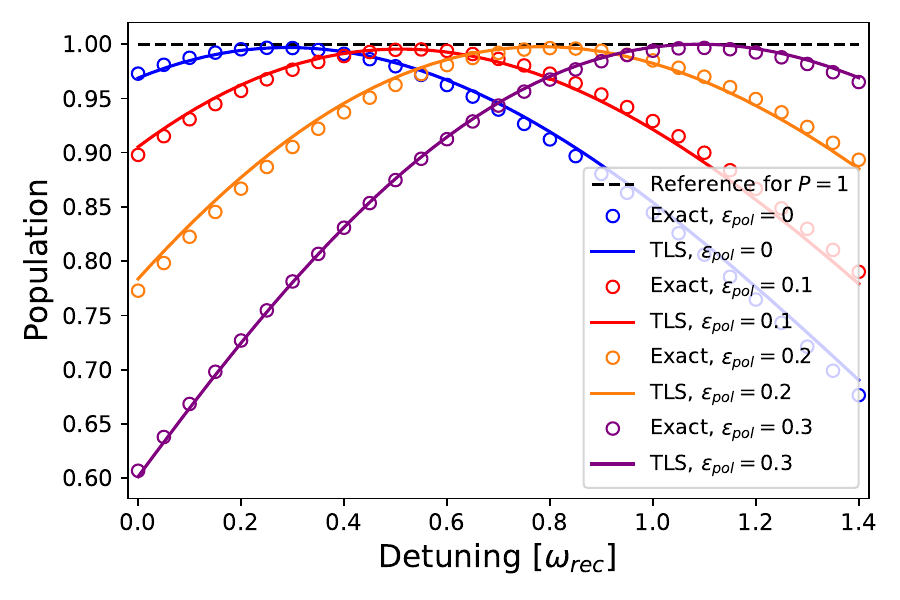}} 
    \caption{(a) Population in state $|1\rangle$ after double Bragg Gaussian-pulse under different polarization errors as a function of pulse width $\tau$, with fixed parameters $\Delta=0,\, \Omega_R=2\,\omega_{rec}$. (b) Population in state $|1\rangle$ with a fixed Gaussian-pulse width and peak Rabi frequency ($\tau=0.47\,\omega_{rec}^{-1}$, $\Omega_R=2.0\,\omega_{rec}$) in the first Rabi cycle as a function of detuning $\Delta$ under different polarization errors $\epsilon_{pol}=(0.0,\,0.1,\,0.2, \,0.3)$ in (blue, red, orange, purple). The solid lines are predictions of the effective TLS Hamiltonian (\ref{eq:H_eff}) and circles are the exact numerical solution of the full Hamiltonian (\ref{eq:1}).}
    \label{Fig. pol_err_compare}
 \end{figure}

In Fig.~\ref{Fig. pol_err_compare}a, we quantitatively compare the dynamics predicted by the pulse-area condition~(\ref{eq:pulse-area}), the effective TLS Hamiltonian~(\ref{eq:H_eff}) and the exact numerical solution of the real-space Hamiltonian~(\ref{eq:1}) for a fixed peak Rabi frequency $\Omega_R =2\, \omega_{rec}$. We observe that the first peak population decreases with increasing polarization errors in the system. Moreover, the pulse-area formula~\eqref{eq:pulse-area}, as shown by the black line in Fig.~\ref{Fig. pol_err_compare}a, inaccurately captures the true dynamics of the DBD without polarization error in the first Rabi cycle, as shown by the exact numerical solution (blue circles in Fig.~\ref{Fig. pol_err_compare}a). In contrast, our effective TLS Hamiltonian~(\ref{eq:H_eff}) accurately captures the dynamics within the first Rabi cycle, even under moderate polarization errors up to $\epsilon_{pol}=0.2$. 
\subsubsection{Mitigation with constant detuning control}
To mitigate polarization errors, we proceed to detuning control and explore its effect on the target state population for various polarization errors. In Fig.~\ref{Fig. pol_err_compare}b, we plot the population in state $|1\rangle$ as a function of detuning for different polarization errors with a fixed Gaussian-pulse width and peak Rabi frequency. We find an optimal constant detuning $\Delta_{opt}$ exists, enabling full population inversion for each polarization error. For instance, $\Delta_{opt}/\omega_{rec} =(0.25,\,0.55,\,0.80,\,1.10)$ for $\epsilon_{pol}=(0,\,0.1,\,0.2,\,0.3)$. Considering that typical polarization errors in a retro-reflective setup for DBD experiments can be reduced to below $\epsilon_{pol}=10\%$, it means one can apply a constant detuning up to $0.55\,\omega_{rec}$ (for $\tau=0.47\,\omega_{rec}^{-1}$, $\Omega_R=2.0\,\omega_{rec}$) to mitigate the effect of both AC-Stark shift and polarization errors, provided that the polarization error $\epsilon_{pol}$ is known and fixed throughout the experiment.
\subsubsection{Mitigation with linear detuning control}
 However, if the polarization error in the DBD experiment is unknown or fluctuates from shot-to-shot, a robust protocol is needed to mitigate a range of polarization errors typically encountered in the experiment. We claim that this can indeed be achieved via a time-dependent detuning. The basic idea is that if one can design a time-dependent detuning function $\Delta(t)$, which adiabatically sweeps through the light-shifted resonance during the pulse, one would expect a full population inversion from the adiabatic passage theory~\cite{Steck, Landau1932, Zener1932, Rubbmark1981}. 
 
 With the physical insights gained from the two-level description of DBD, we are able to engineer a linear detuning sweep that exhibits robustness against a range of polarization errors given by
 \begin{align}
       \Delta(t)/\omega_{rec}=\frac{1}{2.5\tau}(t-t_0+\tau), \label{eq:linear_sweep}
 \end{align}
 where $t_0$ and $\tau$ represent the center and width of the Gaussian pulse in Eq.~(\ref{eq:gaussian}), respectively. 
   \begin{figure}[h]
  \centering
  \includegraphics[width=1.0\columnwidth]{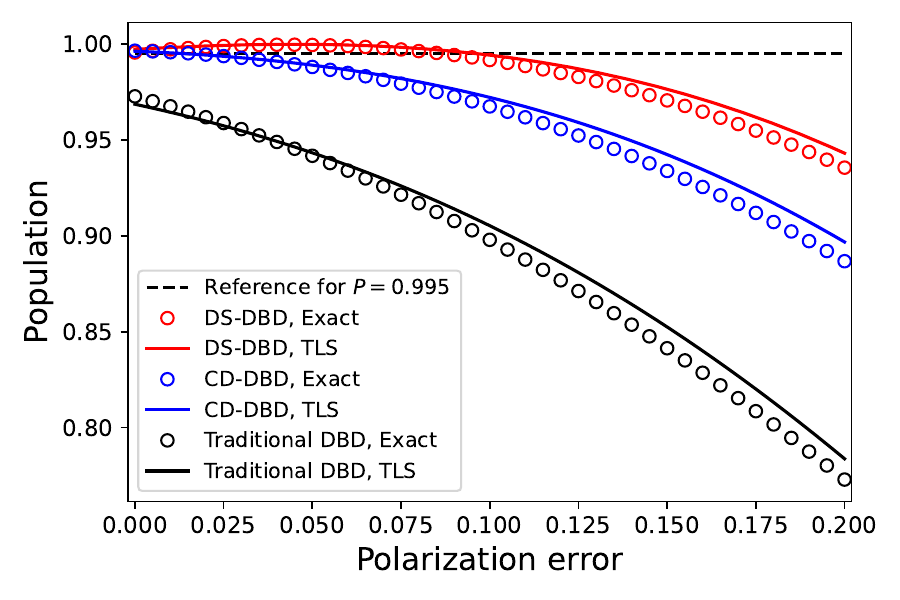}
    \caption{Comparison of double Bragg efficiencies (peak population in state $|1\rangle$ within the first Rabi cycle) as a function of the polarization error among three different protocols: the DS-DBD, the CD-DBD and the traditional DBD. The red curve and circles stand for the TLS dynamics and exact dynamics of the DS-DBD with $\Delta(t)/\omega_{rec}=(t-t_0+\tau) / (2.5\,\tau)$, the blue curve and circles depict the TLS dynamics and exact dynamics of the CD-DBD with $\Delta(t)/\omega_{rec}=0.25$, and the black curve and circles represent the TLS dynamics and exact dynamics of the traditional DBD with no detuning, i.e., $\Delta(t)=0$. The dashed black line stands for the threshold efficiency $P=0.995$. The Gaussian pulse parameters used are $\Omega_R=2.0\,\omega_{rec}$ and $\tau=0.47\,\omega_{rec}^{-1}$.}
    \label{Fig:detuning sweep}
 \end{figure}
 In Fig.~\ref{Fig:detuning sweep}, we compare the detuning-sweep DBD (DS-DBD) protocol against the traditional DBD protocol without detuning and the constant detuning mitigated DBD (CD-DBD) protocol with $\Delta =0.25\, \omega_{rec}$. A significant improvement in the maximum double Bragg efficiency within the first Rabi cycle is observed for DS-DBD compared to the other two protocols across a broad range of polarization errors: $\epsilon_{pol}\in[0,\,0.2]$. In particular, the linear sweep defined by Eq.~(\ref{eq:linear_sweep}) demonstrates robustness against up to $8.5\%$ polarization errors while maintaining a double Bragg efficiency exceeding $99.5\%$. The highest efficiency achieved by DS-DBD in this range is $99.976\%$ at $\epsilon_{pol}=0.045$.
\subsubsection{Mitigation with OCT detuning control}\label{section_OCT_pol_err}

In order to achieve even higher efficiencies, we employ optimal control theory for a generalized detuning control. The optimal control theory that we will use is based on the optimizer in Q-CTRL's Boulder Opal package~\cite{QCtrl}. As the first OCT example, we optimize the efficiency of the BS and its robustness against polarization errors. We consider the DBD with a Gaussian pulse (Eq.~\eqref{eq:gaussian}) driving the interaction and a time-dependent detuning $\Delta(t)$ constrained to have a maximum value of $4\,\omega_{rec}$. Therefore, the optimization variables will be the peak Rabi frequency, the width of the Gaussian, the center of the Gaussian, and the time-dependent detuning: $\left(\Omega_R,\, \tau,\, t_0, \,\Delta(t)\right)$. Details of the optimization, cost function, and optimal variables, including the time-dependent detuning, can be found in App.~\ref{App:QCTRL}.

To derive the effective TLS Hamiltonian (Eq.~\eqref{eq:H_eff}), it was assumed that the detuning $\Delta$ is constant, which does not apply for a general time-dependent detuning $\Delta(t)$. Therefore, OCT optimizations will be based on the 5-level Hamiltonian of Eq.~\eqref{eq: H_full}.

 \begin{figure}[h]
  \centering
  \includegraphics[width=\columnwidth]{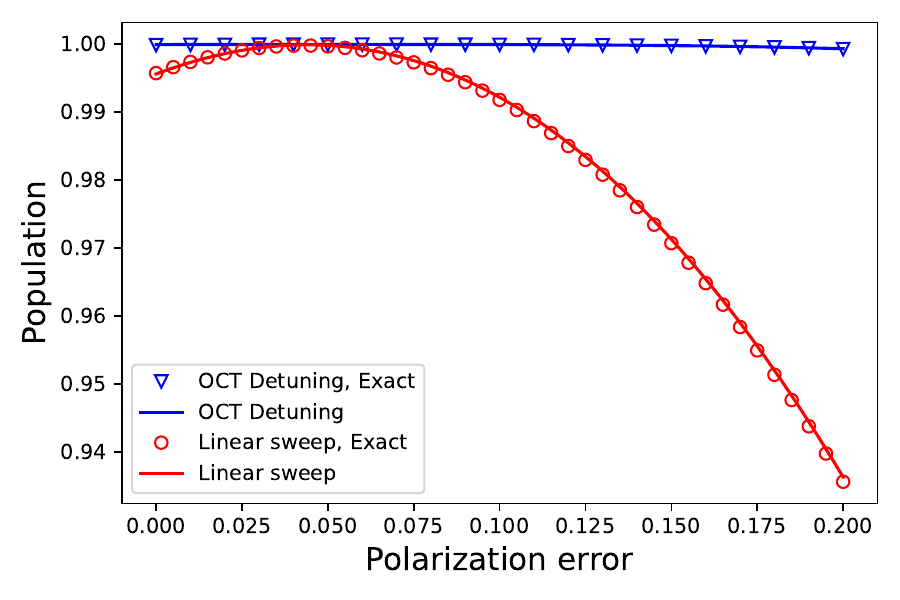}
    \caption{Comparison of the population in the target state after a Gaussian BS pulse for the linear sweep $\Delta(t)= (t-t_0+\tau)/ (2.5\,\tau)$, with $(\Omega_R,\, \tau,\, t_0) = (2\,\omega_{rec}, 0.47\,\omega_{rec}^{-1}, 0)$ (red), including the exact numerical simulation (red circles) and the optimized time-dependent detuning given in App.~\ref{App:QCTRL} with $\left(\Omega_R,\, \tau, \,t_0\right) = (1.264\,\omega_{rec},\, 0.915\,\omega_{rec}^{-1}, \,4.065\,\omega_{rec}^{-1})$ (blue), including the exact numerical simulation (blue triangles).}
    \label{Fig:PopMomentum0}
 \end{figure}
The population in the target state
after applying a Gaussian BS pulse is shown in Fig.~\ref{Fig:PopMomentum0}. This figure compares the results of the linear sweep described by Eq.~\eqref{eq:linear_sweep}, with those obtained using the optimized time-dependent detuning given in Fig.~\ref{Fig:TDetuning}a in App.~\ref{App:QCTRL}. Since the target state population remains flat around unity under the OCT detuning control, we need to identify a more suitable metric to compare the two time-dependent detuning protocols. To achieve this, we compute the average distance of the population from unity over the polarization error range $\epsilon_{pol} \in [0,\,0.1]$, representing feasible values for the experiments. In this region, for exact numerical calculations in position space, the average distance of the population from unity is $1.2 \times 10^{-4}$ under the OCT detuning control, implying that the population averages above $99.988\%$. In contrast, the average population is $99.85\%$ using a linear sweep. Therefore, the OCT protocol demonstrates, on average, more than one order of magnitude improvement over the linear sweep protocol in the region $\epsilon_{pol} \in [0,\,0.1]$. Moreover, exact numerical calculations in position space indicate a population transfer above $99.95\%$ for polarization errors up to $17\%$. Thus, this OCT detuning control allows us to almost completely compensate for unknown polarization errors over the optimized range. Once the polarization errors for a given experiment are characterized, even more realistic distributions can be considered in the optimization.

 \section{Doppler effects and mitigations} \label{IV}
 
\subsection{Losses and asymmetry due to Doppler effects}
\begin{figure}[h]
  \centering
  \includegraphics[width=1.0\columnwidth]{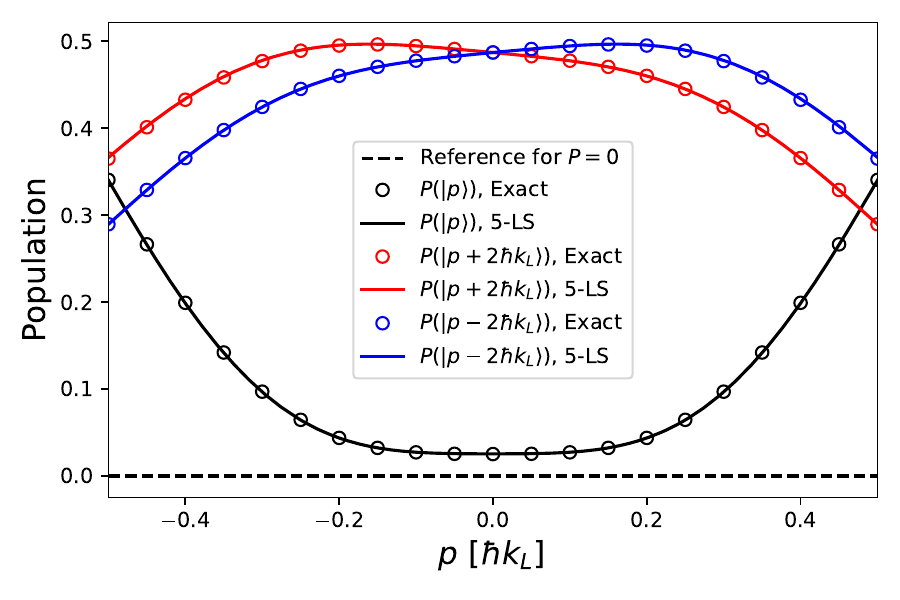}
    \caption{Population in momentum states $\{|p\rangle,\,|p+2\hbar k_L\rangle,\,|p-2\hbar k_L\rangle\}$ after a Gaussian BS-pulse at classical resonance ($\Delta=0$) as a function of the initial momentum $p$. The Gaussian-pulse parameters used are $\Omega_R=2\omega_{rec},\,\tau=0.45\omega_{rec}^{-1}$. The initial momentum width used in the exact numerical calculation is $\sigma_p=0.01\,\hbar k_L$.}
    \label{Fig:DOPPLER}
 \end{figure}

To study Doppler effects on the double Bragg diffraction, we apply the five-level Hamiltonian theory (denoted as 5-LS in the figure legends) to the Gaussian-pulse DBD of an initial momentum eigenstate $|p\rangle$. In Fig.~\ref{Fig:DOPPLER}, the results of the five-level Hamiltonian in the interaction picture (see App.~\ref{A}) and the exact numerical solutions are shown.
The populations in the bare momentum states $\{|p\rangle,\,|p+2\hbar k_L\rangle,\,|p-2\hbar k_L\rangle\}$ after a Gaussian BS-pulse at the traditional resonance condition ($\Delta=0$) are compared. The range of momentum $p$ considered in Fig.~\ref{Fig:DOPPLER} is larger than the typical momentum width $\sigma_p=\sqrt{mk_B T}$ prepared for an AI experiment using delta-kick-cooled (DKC) ultracold atoms as the resource, whose values are well below $0.1\,\hbar k_L$\cite{DKC-Chu-1986, DKC-Ammann-1997, DKC-Morinaga-1999, Müntinga-2013, Kovachy-PRL-2015, Deppner-PRL-2021}. Fig.~\ref{Fig:DOPPLER} shows that the effective $5$-level Hamiltonian accurately captures the dynamics of the full Hamiltonian and correctly predicts the losses due to the AC-Stark shift and the momentum selectivity. In DBD, the momentum selectivity is a phenomenon where, as the initial momentum $p$ deviates further from the resonant value ($p=0$), the transition probability into $|p\pm2\hbar k_L\rangle$ will decrease, as shown in Fig.~\ref{Fig:DOPPLER}. See App.~\ref{C} for a detailed example of momentum selectivity in DBD.

\begin{figure}[h]
  \centering
  \includegraphics[width=1.0\columnwidth]{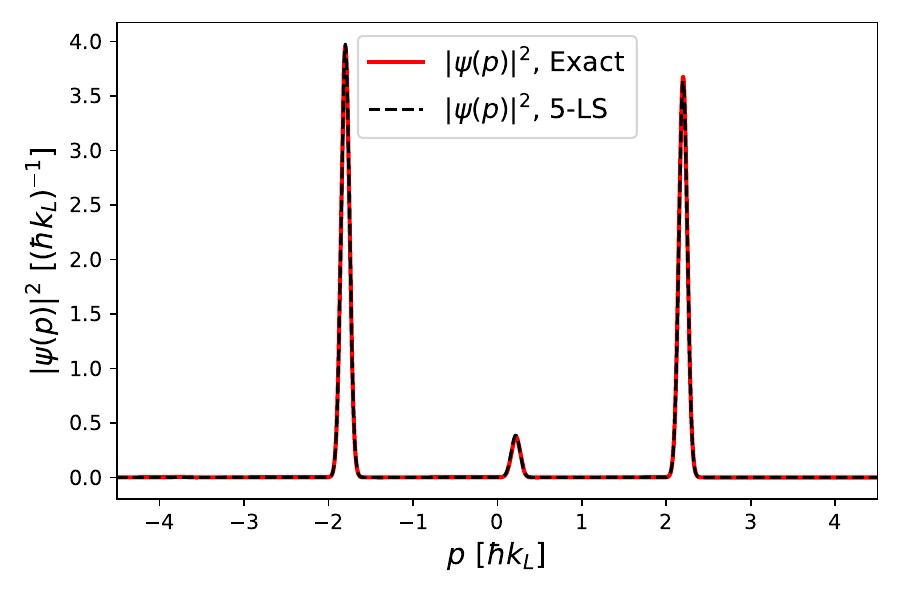}
    \caption{Final wave packet in momentum space $|\psi(p)|^2$ after a BS-pulse showing losses and asymmetry due to Doppler effects with an initial momentum width $\sigma_p= 0.05\,\hbar k_L$ and COM momentum $p_0=0.2\,\hbar k_L$. The results of the 5-level theory and the exact numerical solution are indistinguishable in the plot. Here, the $-2\hbar k_L$ transition is preferred over the $+2\hbar k_L$ transition due to $p_0>0$ and the pulse parameters used are $\Omega_R=2\omega_{rec},\,\tau=0.45\omega_{rec}^{-1}$ with $\Delta=0$.}
    \label{Fig: asymmetry}
\end{figure}

Another important consequence of the Doppler detuning is the asymmetry in the populations between the left and right Bragg diffractions observed for small initial momenta near $p=0$. This asymmetry is characterized by the slope of $P(|p+2\hbar k_L\rangle)$ at $p=0$ after the BS-pulse. Intuitively, this can be understood by noting that the differential light shift $\delta_{diff}=- \Delta - 3 \Omega^2/(64\,\omega_{rec})$, derived in Sec.~\ref{III} for the $p=0$ case, is negative at the traditional resonance condition ($\Delta=0$). The quadratic dispersion relation $E(p)=p^2 / (2m)$ 
gives a negative (positive) differential shift for the left (right) diffraction with a finite positive initial momentum $p_0>0$, as follows:
\begin{align}
    \Delta E|_{L}&=E(-2\hbar k_L+p_0)-E(p_0)-\big(E(-2\hbar k_L)-E(0)\big)\nonumber\\&=-\frac{2\hbar k_L}{m}p_0,\\    
    \Delta E|_{R}&=E(2\hbar k_L+p_0)-E(p_0)-\big(E(2\hbar k_L)-E(0)\big)\nonumber\\&=\frac{2\hbar k_L}{m}p_0,
\end{align}
 which approaches (deviates further from) the light-shifted double-Bragg resonance. Therefore, for a positive initial momentum $p_0>0$, the $-2\hbar k_L$ transition is preferred over the $+2\hbar k_L$ transition, as shown in Fig.~\ref{Fig: asymmetry}. 
 There are two physical consequences due to this asymmetry: (a) An incoming Gaussian wave packet with a COM momentum $p_0=0$ will be split, after the DBD, into two symmetrically distorted Gaussian wave packets in the two outgoing ports with their mean momenta slightly shifted towards zero. (b) An incoming Gaussian wave packet with a COM momentum $p_0\neq0$ will, after the DBD, have imbalanced populations in the two outgoing ports, depending on the sign of $p_0$ (see Fig.~\ref{Fig: asymmetry}). 
 \subsection{Doppler effect mitigation via constant or linear detuning control}\label{subsection:LinearSweep}
\begin{figure}[h]
  \centering
  \subfigure[]{\includegraphics[width=1.0\columnwidth]{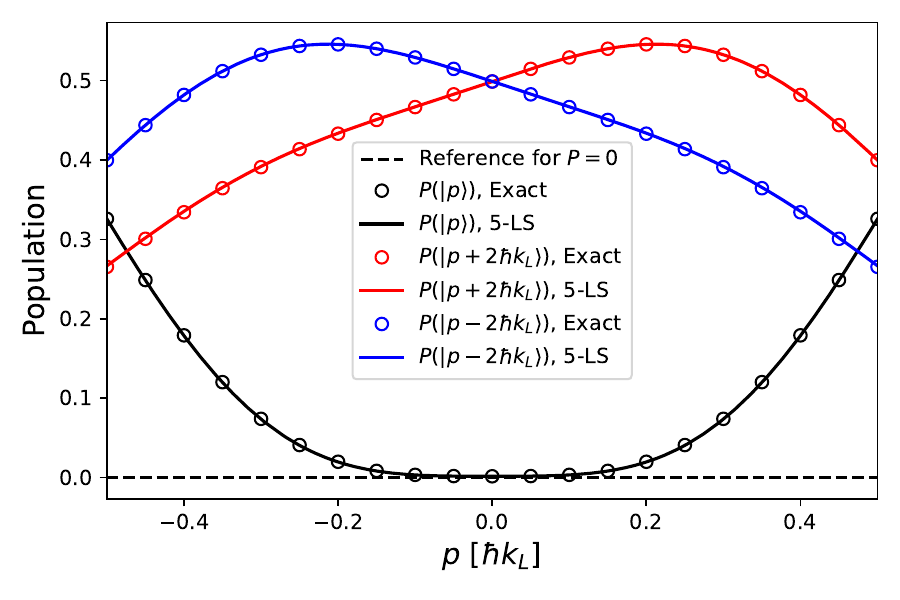}} 
  \subfigure[]{\includegraphics[width=1.0\columnwidth]{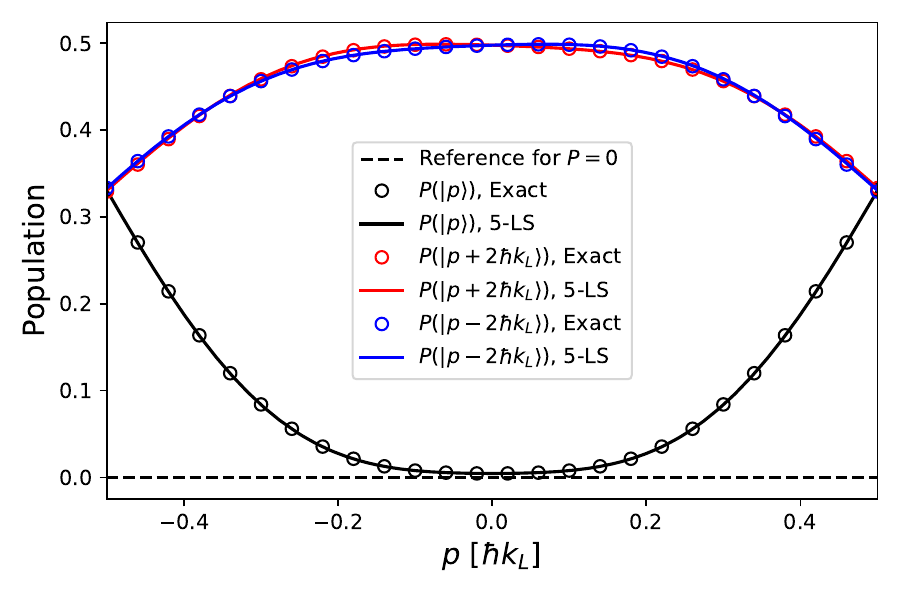}} 
    \caption{Population in different momentum states after a Gaussian BS pulse as a function of the initial momentum $p$ for (a) a constant detuning control $\Delta/\omega_{rec}=0.345$ and (b) a linear detuning sweep $\Delta(t)/\omega_{rec}= (t+0.9\,\tau) / (5\,\tau)$. The Gaussian-pulse parameters used in both cases are $\Omega_R=2\,\omega_{rec},\,\tau=0.45\,\omega_{rec}^{-1}$. The initial momentum width used in the exact numerical calculation is $\sigma_p=0.01\,\hbar k_L$.}
    \label{Fig: Doppler_const_detuning}
 \end{figure}

 To mitigate the detrimental Doppler effects on the efficiency of double Bragg beam-splitters, one can either apply a constant detuning to completely eliminate losses around $p=0$ but at the cost of a more severe asymmetry (see Fig.~\ref{Fig: Doppler_const_detuning}a for a particular constant detuning mitigation), or apply a linear detuning sweep to eliminate the asymmetry while still allowing for some finite losses compared to the constant detuning control (see Fig.~\ref{Fig: Doppler_const_detuning}b for a particular linear detuning sweep mitigation). 
In the next subsection, we will demonstrate how it is possible to eliminate both the losses and the asymmetry by employing a generalized time-dependent detuning optimized via OCT.
\subsection{Doppler effect mitigation via OCT}\label{section_OCT_Doppler_only}
Similar to the optimization against polarization errors in Sec.~\ref{section_OCT_pol_err}, the optimization against Doppler effects involves sampling a uniform distribution of the initial momentum $p \in [-0.3\hbar k_L, 0.3\hbar k_L]$ and minimizing both the deviation from a $50/50$ population distribution between the states $\ket{p+2\hbar k_L}$ and $\ket{p-2\hbar k_L}$, as well as minimizing the asymmetry. Details of the cost function and optimization results can be found in App.~\ref{App:QCTRL}.

\begin{figure}[h]
  \centering
\includegraphics[width=1.0\columnwidth]{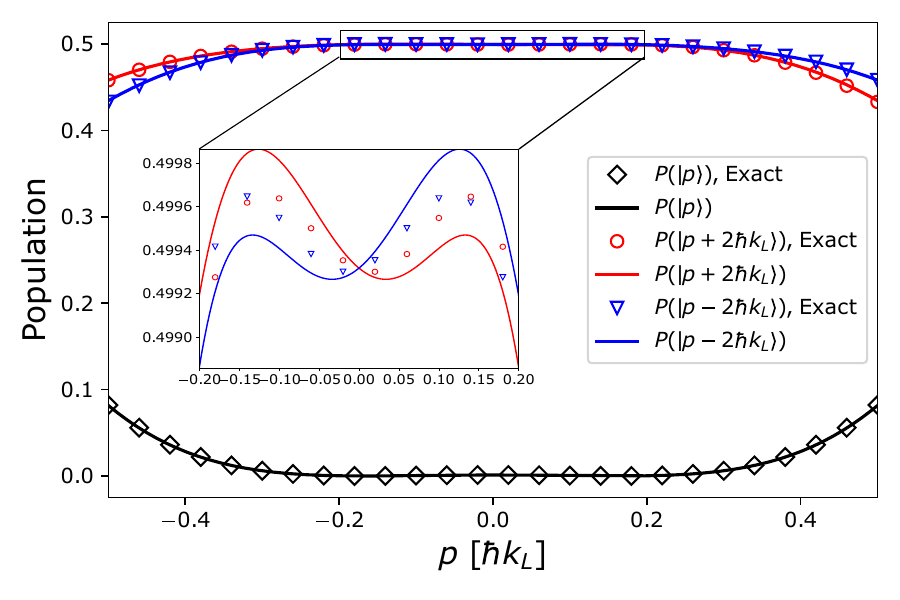}
    \caption{Population in different momentum states after a Gaussian BS pulse as a function of the initial momentum $p$ in different momentum states for the optimized time-dependent detuning shown in Fig.~\ref{Fig:TDetuning}b in App.~\ref{App:QCTRL}. The Gaussian-pulse parameters used are $\left(\Omega_R,\, \tau,\, t_0\right) = (2.079\,\omega_{rec},\, 0.534\,\omega_{rec}^{-1}, \,2.463\,\omega_{rec}^{-1})$. Inset: A zoom-in for the range $p\in [-0.2\,\hbar k_L,\, 0.2\,\hbar k_L]$, excluding the population in the initial state. }
    \label{Fig:PopPolarization0}
\end{figure}
In Fig.~\ref{Fig:PopPolarization0}, we plot the results of the final populations in the bare momentum states $\ket{p}$, $\ket{p+2\hbar k_L}$ and $\ket{p-2\hbar k_L}$ as a function of the initial momentum $p$ using the optimal time-dependent detuning given by the OCT and find that the latter two are closer to $0.5$ than the previous simple detuning control results. The inset plot in Fig.~\ref{Fig:PopPolarization0} is a zoom-in in the range $p\in [-0.2\,\hbar k_L, 0.2\,\hbar k_L]$ to take a closer look at the almost flat behavior of the populations in $\ket{p+2\hbar k_L}$ and $\ket{p-2\hbar k_L}$. With this zoom-in, considering that the $y-$axis only spans over a range of $0.1\%$ in the population, one notices that the asymmetry is successfully mitigated according to the exact numerical solutions. At this order of magnitude, the small deviation between the exact numerical calculations and the OCT simulations is probably due to the fact that the exact numerical solutions assume a $0.01\,\hbar k_L$ momentum width for the initial wave packet whereas OCT assumes an infinitely narrow momentum distribution. Another reason for the deviation could be that higher momentum states beyond the five-level description are populated during the Gaussian BS pulse. Furthermore, we see a considerable improvement in the populations of the states $\ket{p+2\hbar k_L}$ and $\ket{p-2\hbar k_L}$ compared to the linear detuning sweep shown in Fig.~\ref{Fig: Doppler_const_detuning}b, not only for the boundary $p$ values but also for a larger range of intermediate $p$ values, making it less momentum-selective.

\subsection{Combined mitigation via OCT}

To achieve robustness against both polarization errors and Doppler effects, we sample from a uniform distribution of polarization errors $\epsilon_{pol}\in [0, 0.1]$ (as detailed in Sec.~\ref{section_OCT_pol_err}) and a uniform distribution of initial momenta $p \in [-0.3\hbar k_L, 0.3\hbar k_L]$ (as outlined in Sec.~\ref{section_OCT_Doppler_only}). We then minimize the deviation from a $50/50$ population distribution between the states $\ket{p+2\hbar k_L}$ and $\ket{p-2\hbar k_L}$, as well as the asymmetry in their populations. 

\begin{figure}[t]
  \centering
  \subfigure[\,Linear detuning sweep \vspace{-2cm} ]{\includegraphics[width=1.0\columnwidth]{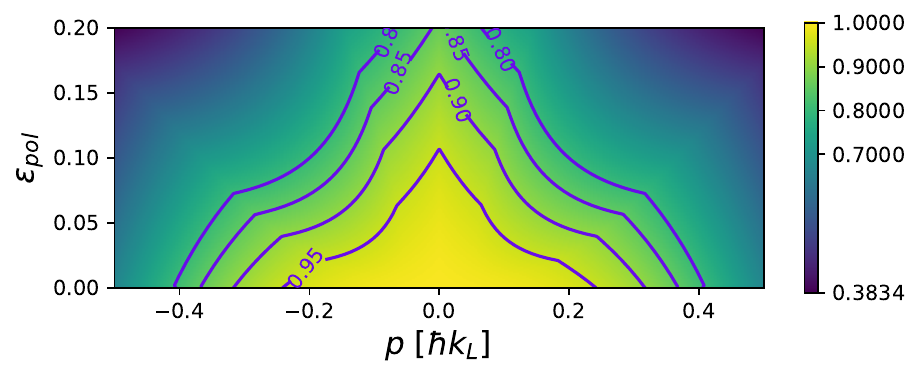}}\\[-0.2cm]
  \subfigure[\,Optimized time-dependent detuning]{\includegraphics[width=1.0\columnwidth]{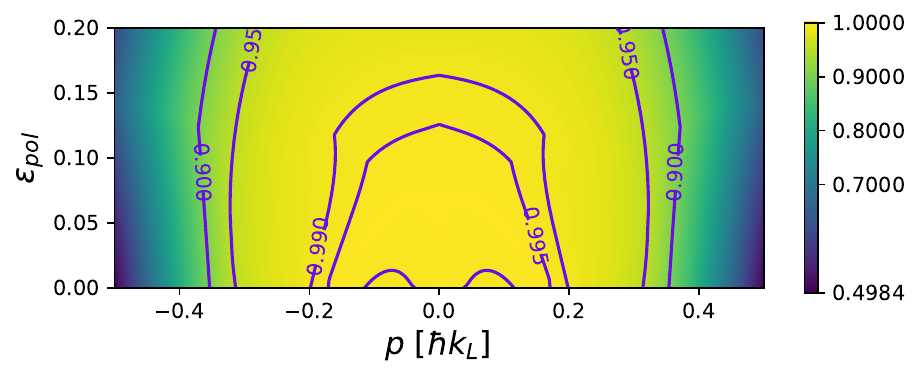}} 
    \caption{OCT beam-splitter efficiency optimized across a range of polarization errors and initial momenta for two protocols: (a) the OCT BS efficiency of a linear detuning sweep $\Delta(t)/\omega_{rec}=\frac{1}{5\tau}(t+0.9\,\tau)$ with parameters $\left(\Omega, \tau, t_0\right) = (2\,\omega_{rec},\, 0.45\,\omega_{rec}^{-1}, 0)$. (b) the performance of the OCT optimized detuning control (given by Fig.~\ref{Fig:TDetuning}c in App.~\ref{App:QCTRL}) with $\left(\Omega,\, \tau,\, t_0\right) = (1.264\,\omega_{rec}, \,0.915\,\omega_{rec}^{-1},\, 4.065\,\omega_{rec}^{-1})$.}
    \label{Fig:3DcontourPlot}
 \end{figure}

In Fig.~\ref{Fig:3DcontourPlot}, we present the OCT beam-splitter efficiency, aimed at accounting for both population transfer and the symmetry between $P(|p-2\hbar k_L\rangle)$ and $P(|p+2\hbar k_L\rangle)$. We refer to this new metric as the \textit{OCT BS efficiency} to distinguish it from the DBD efficiency described in Sec.~\ref{III}. More details about the OCT BS efficiency can be found in App.~\ref{App:QCTRL}. For comparison, we show the OCT BS efficiencies for both linear detuning control (in Fig.~\ref{Fig:3DcontourPlot}a) and OCT optimized time-dependent detuning (in Fig.~\ref{Fig:3DcontourPlot}b). These contour plots demonstrate robustness against both polarization errors and Doppler detuning. 
The linear detuning sweep achieves an OCT BS efficiency of $0.95$ within a triangular region in the $p$-$\epsilon_{pol}$ space, which rapidly decreases as $p$ and $\epsilon_{pol}$ increase. In contrast, the OCT-optimized detuning achieves an OCT BS efficiency of $0.99$ over a square region where the initial momentum ranges up to $p=\pm 0.18\,\hbar k_L$ and polarization errors extend up to $\epsilon_{pol} = 0.12$.

\begin{figure}[h!]
  \centering
  \includegraphics[width=1.0\columnwidth]{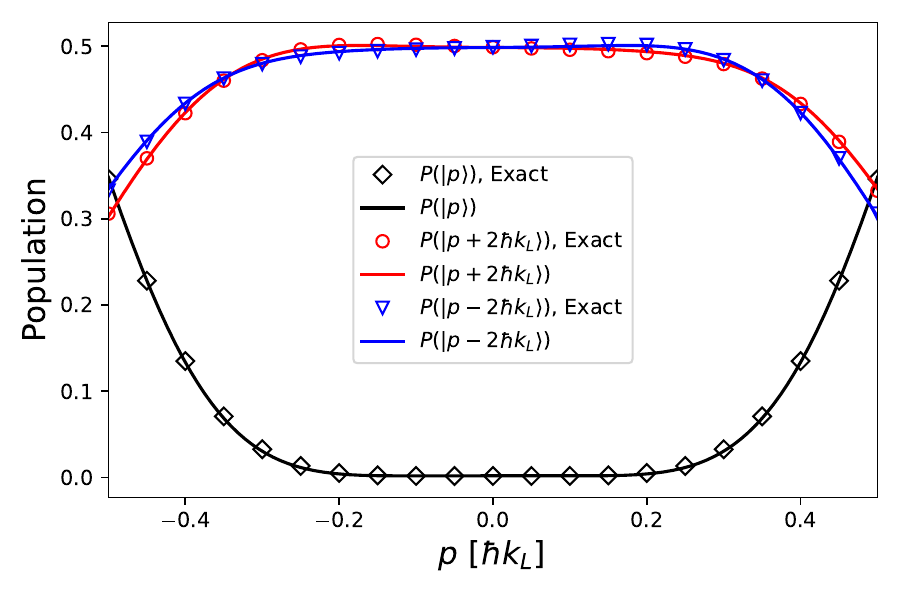}
    \caption{Populations in different momentum states after a Gaussian BS pulse with a polarization error $\epsilon_{pol} = 0.1$ is shown as a function of the initial momentum $p$ with the OCT detuning control detailed in Fig.~\ref{Fig:TDetuning}c in App.~\ref{App:QCTRL}. The Gaussian pulse parameters used are $\left(\Omega, \,\tau, \,t_0\right) = (2.230\,\omega_{rec}, 0.505\,\omega_{rec}^{-1}, 2.970\,\omega_{rec}^{-1})$. }
    \label{Fig:0.1case}
 \end{figure}

To further analyze the OCT BS performance and its robustness to both polarization errors and Doppler effects, we plot the population in different momentum states after a Gaussian BS pulse as a function of initial momentum $p$ for a specific polarization error $\epsilon_{pol} = 0.1$ in Fig.~\ref{Fig:0.1case}, which corresponds to a cut through Fig.~\ref{Fig:3DcontourPlot}b. We note that the populations are well optimized for initial momenta near $p=0$. However, the performance begins to deteriorate for larger initial momentum values compared to the detuning control results shown in Fig.~\ref{Fig:PopPolarization0}, which were only optimized for $\epsilon_{pol} = 0$.

\begin{figure}[t]
  \centering
  \includegraphics[width=1.0\columnwidth]{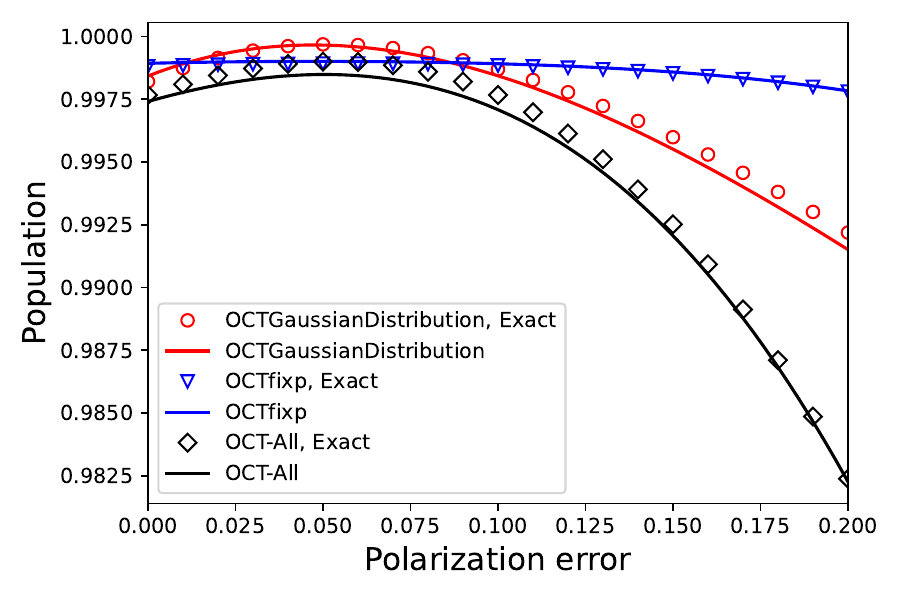}
    \caption{Summed population in $\pm 2\hbar k_L$-ports after a Gaussian BS pulse for an input Gaussian state with a COM momentum $p_0 = 0$ and a finite momentum width $\sigma_p = 0.05\,\hbar k_L$ for three different OCT time-dependent detunings: (a) optimized for a momentum width of $\sigma_p =0.05\,\hbar k_L$ and polarization errors $\epsilon_{pol} \in [0,0.1]$ with pulse parameters $\left(\Omega_R,\, \tau,\, t_0\right) = (1.264\,\omega_{rec}, \,0.915\,\omega_{rec}^{-1}, \,4.065\,\omega_{rec}^{-1})$ (red line); (b) optimized for robustness against polarization errors for $p=0$ with pulse parameters $\left(\Omega_R, \,\tau, \,t_0\right) = (1.617\,\omega_{rec},\, 0.583\,\omega_{rec}^{-1}, 2.859\,\omega_{rec}^{-1})$ (blue line); (c) optimized for robustness against both polarization errors and Doppler effects with pulse parameters $\left(\Omega_R, \,\tau, t_0\right) = (2.079\,\omega_{rec}, 0.534\,\omega_{rec}^{-1}, 2.463\,\omega_{rec}^{-1})$ (black line). We also include the exact numerical solutions with (a) red circles, (b) blue triangles and (c) black diamonds.}
    \label{Fig: ComparisonsModelsFor0.05}
 \end{figure}

Finally, we examine a specific real-world scenario where the input state has a finite momentum width of $\sigma_p = 0.05\,\hbar k_L$. In this case, we apply the same optimization method described earlier in this section, except we opt for Gaussian sampling instead of uniform sampling in the initial momentum $p$. In Fig.~\ref{Fig: ComparisonsModelsFor0.05}, we compare the summed population in $\pm 2\hbar k_L$-ports after the Gaussian BS pulse for three different OCT protocols: (a) OCT protocol optimized for robustness against polarization errors, considering a finite initial momentum width of $\sigma_p = 0.05\,\hbar k_L$ (with detailed detuning shown in Fig.~\ref{Fig:TDetuning}d in App.~\ref{App:QCTRL}); (b) OCT protocol optimized for robustness against polarization errors for $p=0$ (with detailed detuning shown in Fig.~\ref{Fig:TDetuning}a in App.~\ref{App:QCTRL}); (c) OCT protocol optimized for robustness against both polarization errors and Doppler effects (with detailed detuning shown in Fig.~\ref{Fig:TDetuning}c in App.~\ref{App:QCTRL}). As observed, the first protocol (red line and circles in Fig.~\ref{Fig: ComparisonsModelsFor0.05}) specifically optimized for this input momentum width of $\sigma_p = 0.05\,\hbar k_L$ exhibits better overall performance in the range $\epsilon_{pol} \in [0,0.1]$, achieving an average population of $99.92\%$ in the target ports. However, the population obtained from the OCT protocol optimized for $p=0$ is more stable and remains effective for larger polarization errors beyond $0.1$. We also investigated the visible deviations in Fig.~\ref{Fig: ComparisonsModelsFor0.05} between the exact numerical solutions and OCT predictions. These deviations are due to transitions to higher momentum states beyond the five-level description, resulting in occupations of the order between $10^{-5}$ and $10^{-4}$ during the time evolution, which leads to a deviation of the order of $10^{-4}$ in the final population integrated over the target ports.

 \section{Summary and outlook}
 In summary, the double Bragg diffractions under weak-driving conditions can be well-described by analytical models of few-level effective Hamiltonians proposed in the paper. For the ideal atomic cloud at zero temperature with vanishing momentum, a two-level system is enough to capture the essential physics and give rise to the AC-Stark shift for both box-pulse and Gaussian-pulse DBDs with good accuracy compared to the exact numerical solution both with and without polarization error. However, to describe the Doppler effects associated with finite momentum spread of an atomic cloud at a finite temperature and a general time-dependent detuning, a five-level description is needed in order to achieve a good accuracy. We should point out that our theory is neither restricted to the shapes of the pulse or frequency detuning, nor the initial momentum distribution. Moreover, extending our theory to describe higher-order DBDs or sequential DBDs~\cite{Sven-PhD-thesis, Müller-PRL-2008, Chiow-PRL-2011, Kovachy-PRA-2012, Kovachy-Nature-2015} is also straightforward. One only needs to include more levels or account for the relevant sector of the full interaction Hamiltonian (see App.~\ref{A}), together with the AC-Stark shift. With the established effective Hamiltonians, we manage to optimize the DBD efficiency via a constant or linear detuning control with intuitions gained from TLS description and then boost the efficiency with the aid of artificial intelligence based on OCT, whose results are proven to be superior to the optimization results achieved by any human effort. By leveraging the analytical theory developed in this work, along with the support of artificial intelligence, we can design DBD pulses specifically tailored for high-precision AI experiments, ensuring robustness against both finite momentum spread and unavoidable polarization errors.

\section*{acknowledgement}
We thank E. Giese for the insightful discussions and thoughtful comments about this work; we also thank S. Abend for discussions on the feasibility of the time-dependent detunings. The authors gratefully acknowledge financial support from the Deutsche Forschungsgemeinschaft (DFG) CRC 1227 274200144 (DQ-mat) within Project A05, Germany’s Excellence Strategy EXC-2123 QuantumFrontiers 390837967, and through the QuantERA 2021 co-funded Project No. 499225223 (SQUEIS). We also thank the German Space Agency (DLR) for funds provided by the German Federal Ministry for Economic Affairs and Climate Action (BMWK) due to an enactment of the German Bundestag under Grants No. 50WM2450A (QUANTUS-VI), No. 50WM2253A (AI-Quadrat), and No. 50NA2106 (QGYRO+). R. L. acknowledges the usage of LUH’s computer cluster funded by the DFG via Project No. INST 187/742-1 FUGG. N.G. and K.H. acknowledge funding by the AGAPES project - Grant No. 530096754 within the ANR-DFG 2023 Programme.

\textbf{Author contributions:}
N.G. supervised the research project. K.H. and R.L. conceived the idea. R.L. performed the analytical modeling and the exact numerical calculations. V.M.-L. performed the optimizations with OCT. S.S. developed the UATIS Python
Environment for exact numerical calculations with cluster GPU support. R.L. and V.M.-L. drafted the manuscript with the participation of all co-authors.

\clearpage

\begin{appendix}
\section{Full DBD Hamiltonian with Doppler detuning in the interaction picture}\label{A}
Here, we give the full matrix expression of the DBD Hamiltonian~(\ref{eq:1}) without truncation in the basis of symmetric and anti symmetric states with Doppler effects and polarization errors. First, we define the following symmetric and anti-symmetric states with $\pm 2n\hbar k_L$ ($n\geq1$) momentum transfer:
\begin{align}
    \Big\{&|n,+\rangle=\frac{|p+2n\hbar k_L\rangle + |p-2n\hbar k_L\rangle}{\sqrt{2}},\nonumber\\
    &|n,-\rangle=\frac{|p+2n\hbar k_L\rangle - |p-2n\hbar k_L\rangle}{\sqrt{2}}
    \Big\},
\end{align}
where $|n,\pm\rangle$ stands for the symmetric and anti-symmetric state, respectively. We order the basis as 
\begin{align}
    |1\rangle &= |p\rangle,\nonumber\\
    |2\rangle &= |1,+\rangle,\nonumber\\
    |3\rangle &= |1,-\rangle,\nonumber\\
    |4\rangle &= |2,+\rangle,\nonumber\\
    |5\rangle &= |2,-\rangle,\, \cdots
\end{align}
For an atom with an initial momentum $p$ in the central bin $p\in [-\hbar k_L, \hbar k_L)$ and prepared in the momentum eigenstate $|1\rangle=|p\rangle$, it will only evolve in the subspace spanned by the basis $\{|n\rangle\}$ under the DBD Hamiltonian~(\ref{eq:1}). Therefore, one only needs the analytical form for an arbitrary matrix element in order to fully describe the system evolution, and we give all the non-zero matrix elements as below: 
\begin{align}
    &\langle 1|H(t)|1\rangle = \frac{p^2}{2m},\nonumber\\&\langle 1|H(t)|2\rangle = \langle 2|H(t)|1\rangle=\sqrt{2}C(t, \epsilon_{pol})\hbar \Omega(t),\nonumber\\
    &\langle 2n|H(t)|2n\rangle =\langle 2n+1|H(t)|2n+1\rangle = \frac{p^2}{2m} + 4n^2\hbar \omega_{rec},\nonumber\\
    &\langle 2n|H(t)|2n+1\rangle =\langle 2n+1|H(t)|2n\rangle = 4 n \frac{p}{\hbar k_L}\hbar\omega_{rec}  ,\nonumber\\
    &\langle 2n|H(t)|2n+2\rangle =\langle 2n+2|H(t)|2n\rangle = C(t, \epsilon_{pol})\hbar \Omega(t),\nonumber\\
     &\langle 2n+1|H(t)|2n+3\rangle =\langle 2n+3|H(t)|2n+1\rangle \nonumber\\&\qquad\qquad\qquad\qquad\quad= C(t, \epsilon_{pol})\hbar \Omega(t), \label{eq:lab_frame_H_full}
\end{align}
where $n\geq 1$ and $C(t, \epsilon_{pol})=\cos\big[(4\omega_{rec}+\Delta(t))t\big]+\epsilon_{pol}$.
Truncating above matrix up to the basis state $|5\rangle$, one will get the 5-level Hamiltonian as shown in Eq.~(\ref{eq: H_full}). Furthermore, the full DBD Hamiltonian in the laboratory frame with matrix elements given by Eq.~(\ref{eq:lab_frame_H_full}) can be transformed into the interaction picture with respect to the diagonal part $H_0=\text{diag}(\frac{p^2}{2m},\,\frac{p^2}{2m}+ 4\hbar\omega_{rec},\,\frac{p^2}{2m}+ 4\hbar\omega_{rec},\cdots)$, and the resulting interaction-picture Hamiltonian matrix elements are given by
\begin{align}
    \bar{H}_{1,2}(t) &=\sqrt{2}\hbar \Omega(t)C(t, \epsilon_{pol})  e^{-i4\omega_{rec} t}\nonumber\\
    & =\bar{H}_{2,1}^*(t),\nonumber\\
    \bar{H}_{2n,2n+1}(t) &= 4 n \frac{p}{\hbar k_L}\hbar\omega_{rec} = \bar{H}_{2n+1,2n}(t)  ,\nonumber\\
    \bar{H}_{2n,2n+2}(t) &=   \hbar \Omega(t) C(t, \epsilon_{pol}) e^{-i4(2n + 1) \omega_{rec}t}\nonumber\\
    & =\bar{H}_{2n+2, 2n}^*(t) \nonumber\\
    & =\bar{H}_{2n+1, 2n+3}(t) = \bar{H}_{2n+3, 2n+1}^*(t),
\end{align}
and all other matrix elements vanish. In particular,
the truncated Hamiltonian in the laboratory frame shown by Eq. (\ref{eq: H_full}) can be transformed into the interaction picture as
\begin{widetext}
\begin{align}
  &\bar H_{Doppler}^{pol}/\hbar= \nonumber\\&\left(\begin{matrix}
0 &  \sqrt{2}\Omega(t)C(t, \epsilon_{pol})e^{-i 4\omega_{rec}t}  &0&0 & 0\\
\sqrt{2}\Omega(t)C(t, \epsilon_{pol})e^{i 4\omega_{rec}t}   &0& 4\frac{p}{\hbar k_L}\omega_{rec}&\Omega(t)C(t, \epsilon_{pol})e^{-i 12\omega_{rec}t}   & 0\\
 0 &4\frac{p}{\hbar k_L}\omega_{rec}  &0&0&\Omega(t)C(t, \epsilon_{pol})e^{-i 12\omega_{rec}t} \\
 0&\Omega(t)C(t, \epsilon_{pol})e^{i 12\omega_{rec}t} &0&0&8\frac{p}{\hbar k_L}\omega_{rec}\\
 0& 0&\Omega(t)C(t, \epsilon_{pol})e^{i 12\omega_{rec}t} &8\frac{p}{\hbar k_L}\omega_{rec}&0
\end{matrix} \right) \label{eq:H_Doppler_int}
\end{align}
where the transformed bases are related to the laboratory frame bases (Eq. (\ref{Eq: five bases}) ) by $|\bar n\rangle = e^{i\frac{H_0}{\hbar}t} | n\rangle$ ($n=1,2,\dots,5$). 
\end{widetext}

\section{Box-pulse DBD with polarization errors}\label{box_pol}
\begin{figure}[h]
  \centering
  \subfigure[]{\includegraphics[width=1.0\columnwidth]{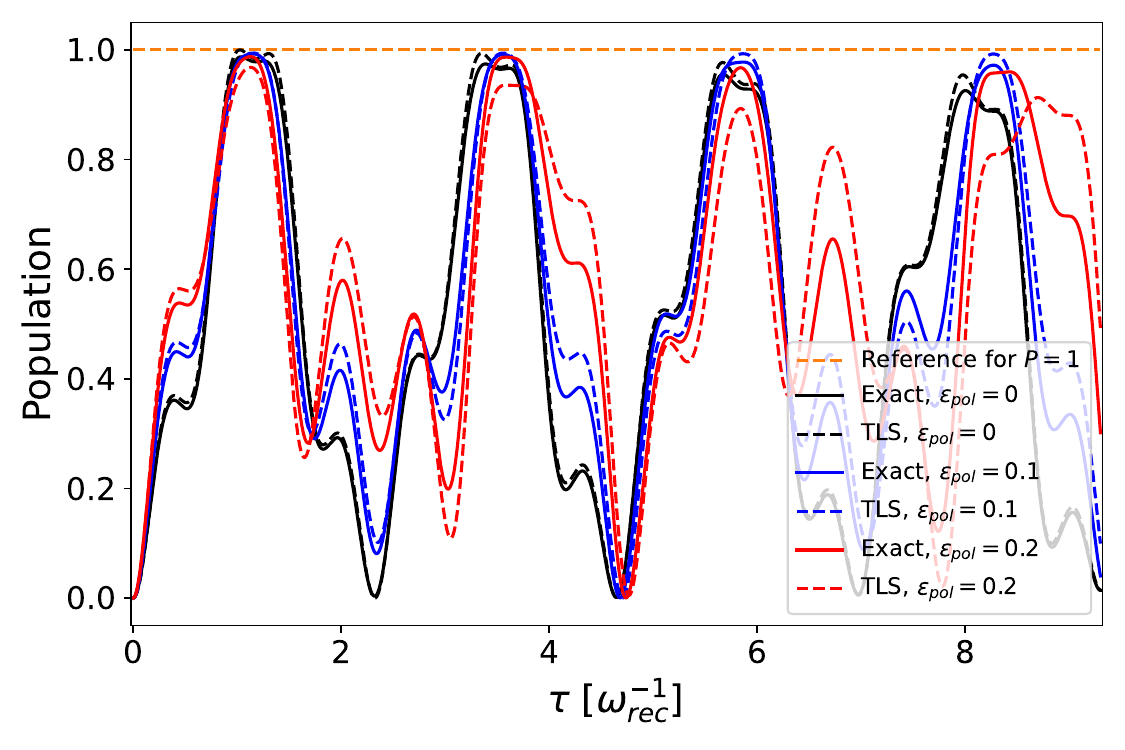}} 
  \subfigure[]{\includegraphics[width=1.0\columnwidth]{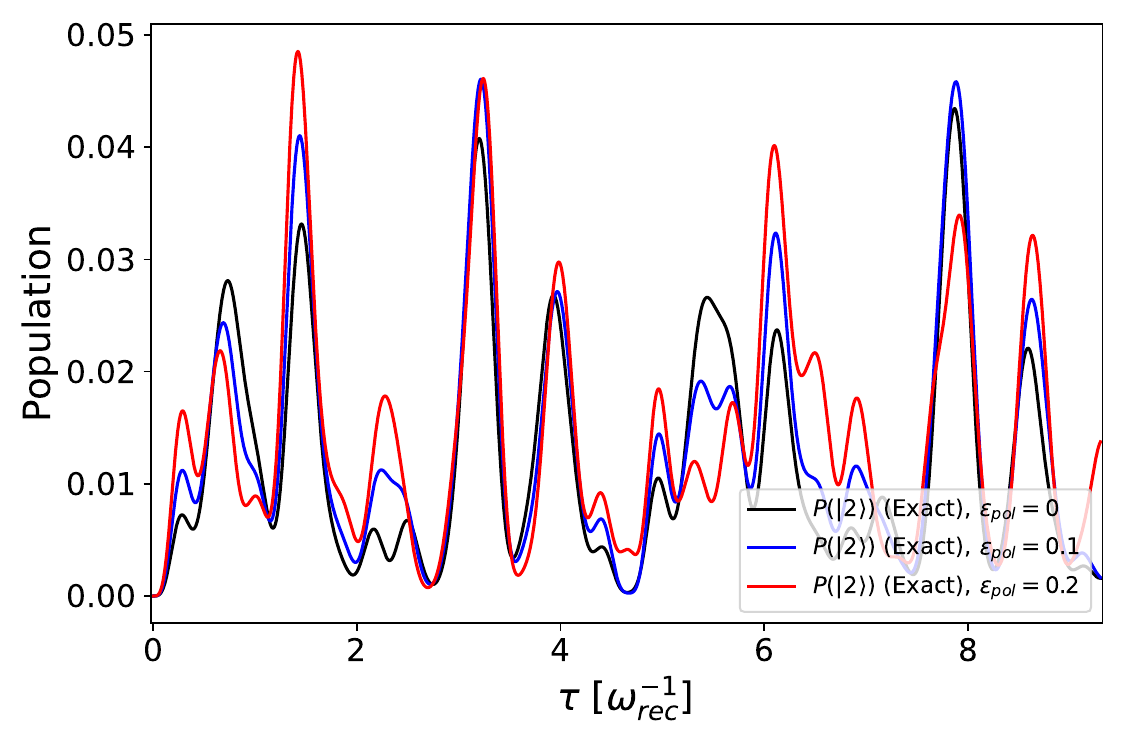}} 
  \caption{Time evolution of population in (a) state $|1\rangle$  and (b) state $|2\rangle$ for a constant Rabi frequency $\Omega=2\omega_{rec}$ and various polarization errors $\epsilon_{pol}=(0,\,0.1,\,0.2)$ with $\Delta=0$. In plot (a), the dynamics predicted by the effective TLS Hamiltonian (\ref{eq:H_eff}) (dashed lines) is compared against the results of the exact numerical solution (solid lines). In plot (b), only the exact numerical results are shown for different polarization errors as solid lines.}
    \label{fig:Rabi_osci_box}
 \end{figure}

For the DBD driven by the time-dependent Rabi frequency given by the box-pulse:
\begin{align}
    \Omega(t)=\Omega\big[\Theta(t)-\Theta(t-\tau)\big],
\end{align}
we numerically solve the effective time-dependent TLS Hamiltonian (\ref{eq:H_eff}) and obtain a 2D parameter scan of $(\Omega,\,\tau)$ and showed in the upper row of Fig. \ref{fig:2D_pol_scan} in the main text which we will study in more details here in the appendix.
\begin{figure}[h]
  \centering
  \includegraphics[width=1.0\columnwidth]{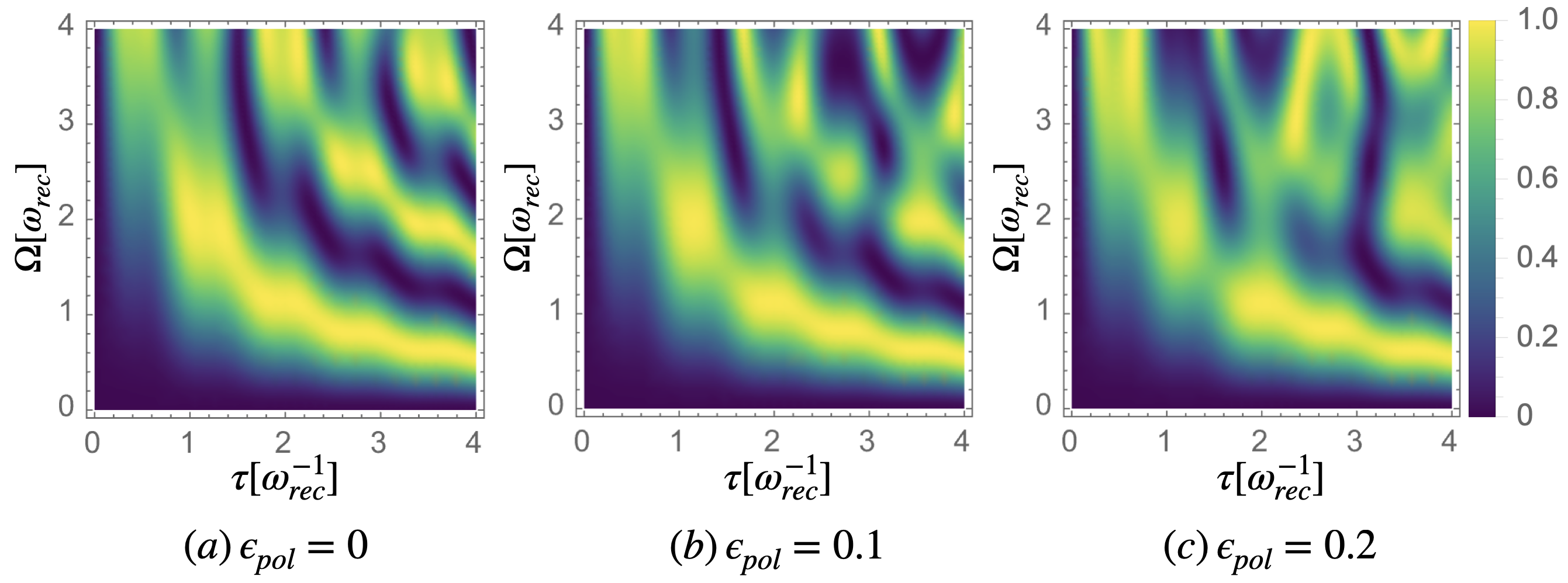}
    \caption{Population in target state $|1\rangle$ after double Bragg box-pulse for various polarization errors $\epsilon_{pol}=(0,\,0.1,\,0.2)$ predicted by the effective TLS Hamiltonian (\ref{eq:H_eff}) as a function of pulse duration and Rabi frequency ($\tau,\,\Omega$) with $\Delta=0$.}
    \label{fig:2D_pol_scan_box}
 \end{figure}
 In Fig.~\ref{fig:2D_pol_scan_box} (a copy of the upper row of  Fig.~\ref{fig:2D_pol_scan}), we observe the pattern of beam-splitter efficiency defined by population in target state $|1\rangle$ after a double Bragg box-pulse gets more and more distorted as polarization error increases from $0$ to $0.2$. If one makes the cut along a fixed Rabi frequency, e.g. $\Omega=2\omega_{rec}$, one will get the time evolution of $P(|1\rangle)$  for different polarization errors shown in Fig.~\ref{fig:Rabi_osci_box} (a).
 Moreover, the deviation of the effective two-level dynamics (dashed lines) from the true dynamics calculated by the exact numerical solution (solid lines) in Fig.~\ref{fig:Rabi_osci_box}(a) can be understood by plotting the time evolution of $P(|2\rangle)$ extracted from the exact numerical solution in Fig.~\ref{fig:Rabi_osci_box}(b). As one can observe, whenever there is a non-negligible fraction of atoms diffracted into higher-order momentum modes, e.g. a peak in $P(|2\rangle)$, there will be a clear deviation of our effective two-level theory to the exact numerical solution simply due to our two-level theory is only valid for describing the first-order DBD up to $\pm 2 \hbar k_L$ transitions. To take into account those higher-order transitions, one in general needs to go back to the full Hamiltonian (\ref{eq:H_int}) and resort to a multi-level description.
  
\section{Momentum selectivity in DBD} \label{C}
\begin{figure}[h]
  \centering
  \subfigure[]{\includegraphics[width=1.0\columnwidth]{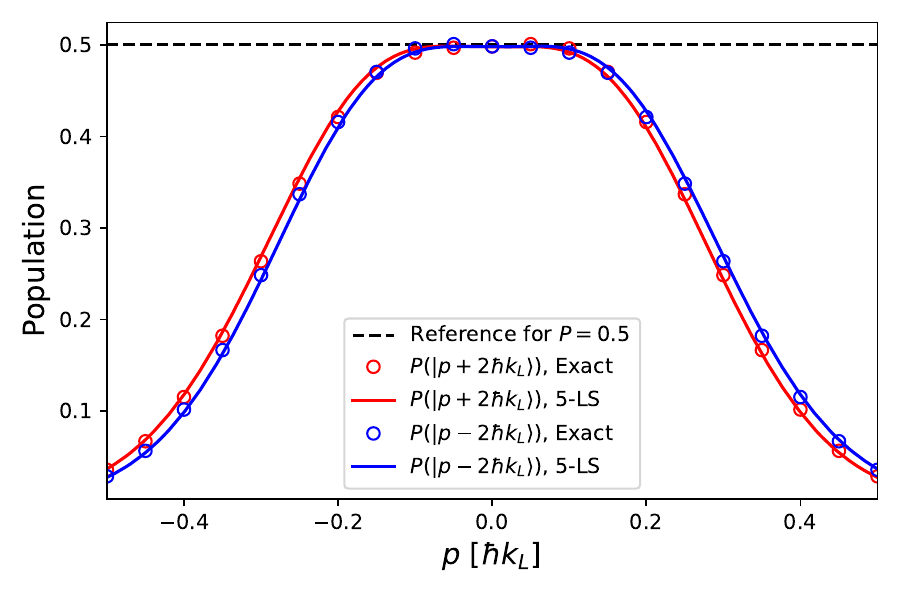}} 
  \subfigure[]{\includegraphics[width=1.0\columnwidth]{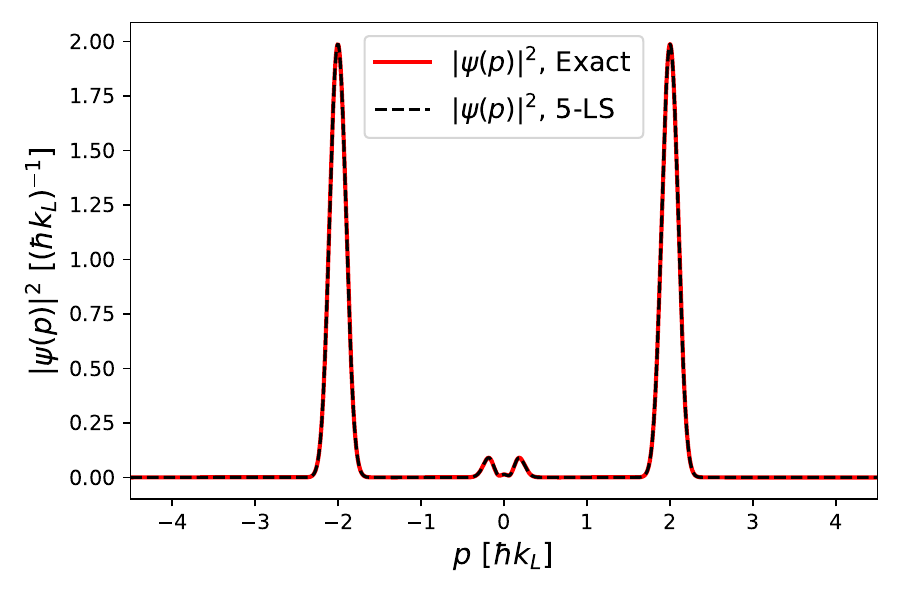}} 
  \caption{(a) Population in different momentum states after a Gaussian BS pulse as a function of initial momentum $p$ with detuning $\Delta=0$. The lines are results of 5-level theory and circles are exact numerical results. The colors red and blue stand for populations in $|p+2\hbar k_L\rangle$ and $|p-2\hbar k_L\rangle$, respectively. (b) Final wave packet in momentum space $|\psi(p)|^2$ after a Gaussian BS pulse showing momentum selectivity due to Doppler effects with an initial momentum width $\sigma_p= 0.1\,\hbar k_L$ and initial COM momentum $p_0=0$. The results of the 5-level theory and the exact numerical solution are indistinguishable in the plot. The Gaussian-pulse parameters are $\Omega_R=1.0\omega_{rec},\,\tau=0.91\omega_{rec}^{-1}$. }
  \label{fig:momentum_selectivity}
\end{figure}

As we have claimed in Sec. \ref{IV} of the main text that the momentum selectivity is a direct consequence of the dynamics governed by the Doppler Hamiltonian (\ref{eq: H_full}) (or its equivalent in the interaction picture given by (\ref{eq:H_Doppler_int})), here we show an example of a Gaussian BS-pulse with a larger width $\tau=0.91\omega_{rec}^{-1}$ (or twice the duration) and the corresponding peak Rabi frequency $\Omega_R=1.0\omega_{rec}$. The momentum acceptance window for this particular BS-pulse is shown in Fig.~\ref{fig:momentum_selectivity}a for the $\pm 2\hbar k_L$ transitions, which is around $(-0.1\hbar k_L,\,0.1\hbar k_L)$. Thus, for an initial wave packet with a momentum width $\sigma_p = 0.1\,\hbar k_L$ comparable to the momentum acceptance window of the BS-pulse, one would expect to observe only the central momentum modes close to $p=0$ get double-diffracted while the higher-momentum tail of the wave packet remains undiffracted as confirmed by the results of both the exact numerical solution and the effective 5-level Doppler Hamiltonian (\ref{eq: H_full}) in Fig.~\ref{fig:momentum_selectivity}b.

\section{OCT optimization}
\label{App:QCTRL}
In this appendix, we show the results of the optimizations, i.e., the time-dependent detunings, and discuss the optimization cost function.

\begin{figure}[h!]
  \centering
  \subfigure[]{\includegraphics[width=1.0\columnwidth]{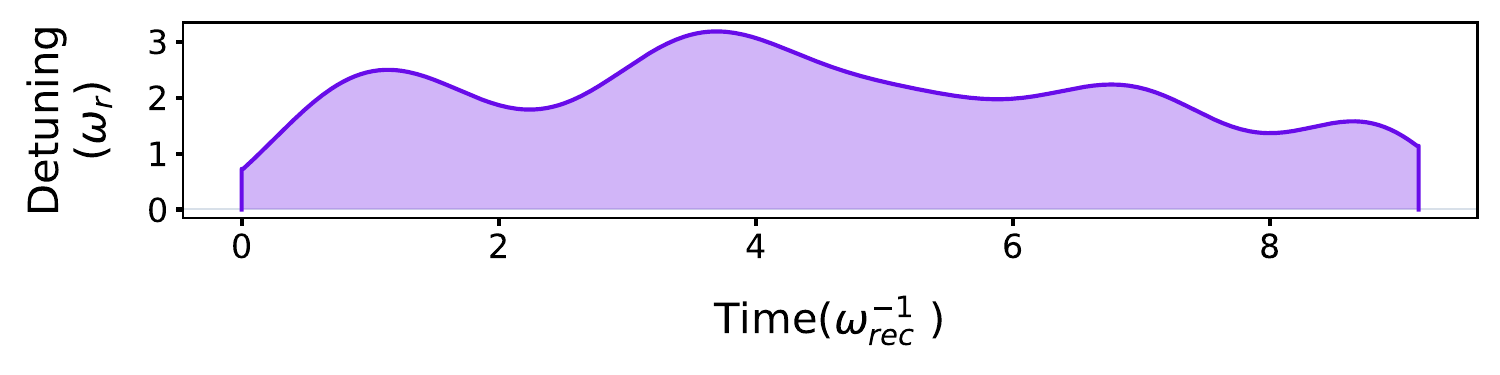}} 
  \subfigure[]{\includegraphics[width=1.0\columnwidth]{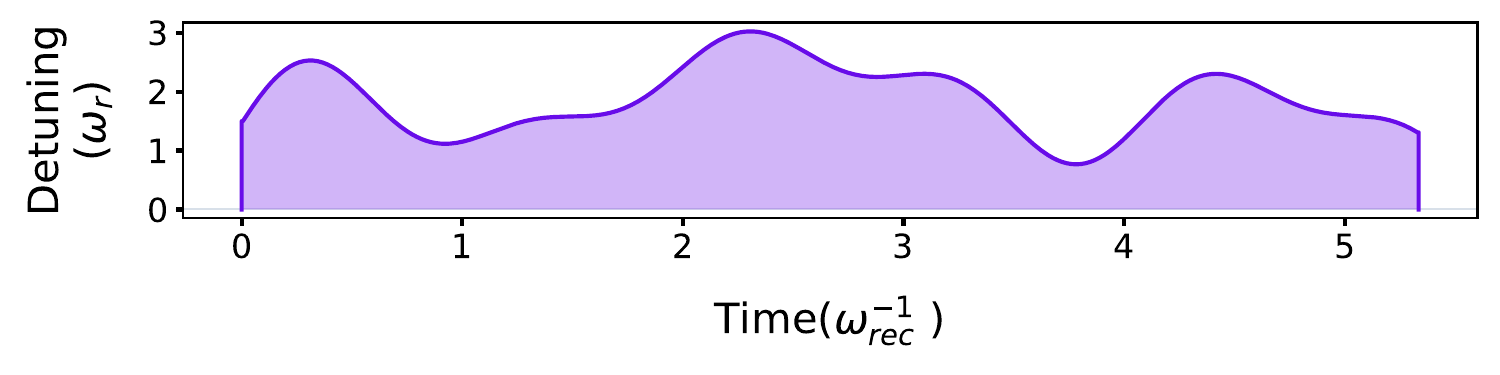}} 
  \subfigure[]{\includegraphics[width=1.0\columnwidth]{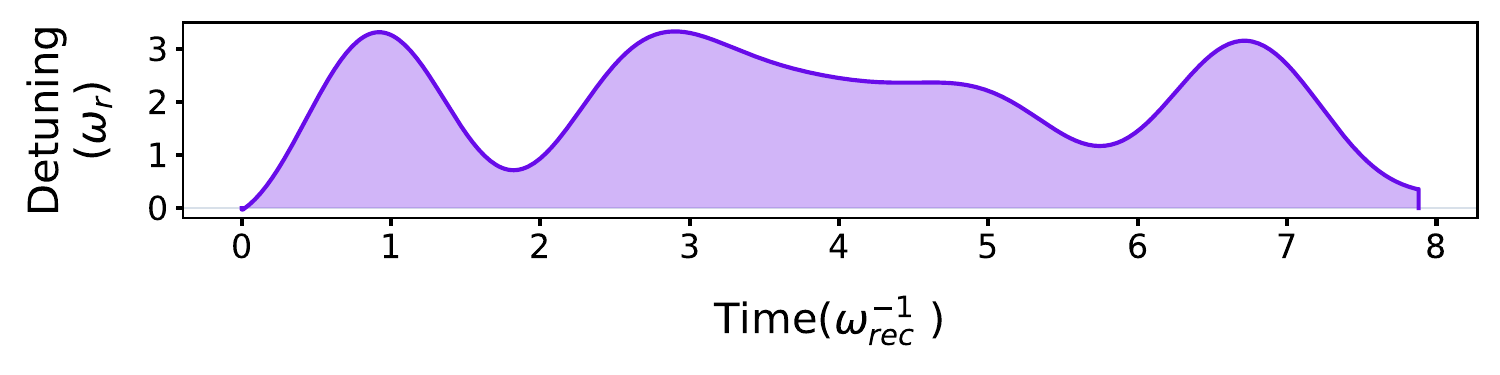}}
  \subfigure[]{\includegraphics[width=1.0\columnwidth]{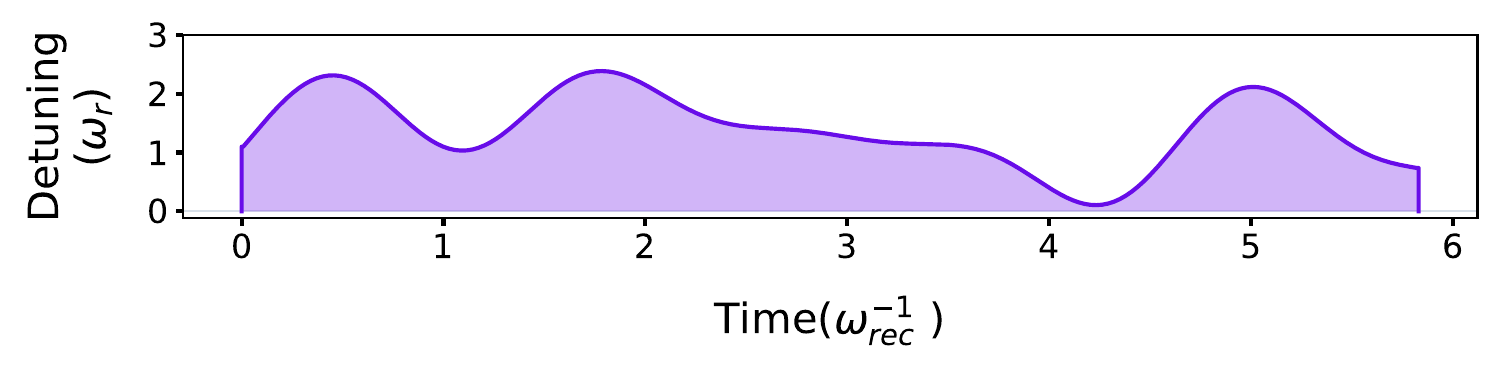}} 
    \caption{Optimal time-dependent detuning to optimize population transfer in different cases. (a) Detuning optimized for robustness against polarization errors with $p=0$. (b) Detuning optimized for robustness against Doppler detuning with $\epsilon_{pol}=0$. (c) Detuning optimized in the presence of both polarization errors and Doppler effects. (d) Detuning optimized for a given momentum width of $\sigma_p= 0.05\,\hbar k_L$ in the presence of polarization errors. All cases are studied using the full 5x5 Hamiltonian description of DBD shown in Eq.~\eqref{eq: H_full}.}
    \label{Fig:TDetuning}
 \end{figure}

To optimize the mitigation of polarization errors in the case of an infinitely narrow initial momentum around $p=0$, we can directly optimize the population transfer to the symmetric state $\left(|p+2\hbar k_L\rangle + |p-2\hbar k_L\rangle \right) / \sqrt{2}$ as there is no cross-term between the symmetric and antisymmetric states. For this case, the optimization results are $\left(\Omega_R, \tau, t_0\right) = (1.264\omega_{rec}, 0.915\omega_{rec}^{-1}, 4.065\omega_{rec}^{-1})$, and the time-dependent detuning is shown in Fig.~\ref{Fig:TDetuning}.

 Even though we used $\left(\Omega, \tau, t_0, \Delta(t)\right)$ as our optimization variables, it is worth highlighting that we obtained similar results if we decide to start with the same pulse parameters as in Sec.~\ref{sec_pol_err_mitigation}, i.e., $(\Omega_R,\, \tau, \,t_0) = (2\omega_{rec}, \,0.47\omega_{rec}^{-1},\, 0)$, and only optimize the time-dependent detuning $\Delta(t)$.
 
In the presence of an initial momentum different from $p = 0$ we have Doppler coupling between the symmetric and antisymmetric states; therefore, we cannot simply optimize the population in the symmetric state. Instead, the cost function for an evolved state after the interaction of duration $t_f$, $|\psi(t= t_f)\rangle_{\epsilon_{pol}, p}$, where the subindices indicate that the evolution occured for a given initial momentum $p$ and polarization error $\epsilon_{pol}$ taken from the range of initial momentum $P = [-0.3\,\hbar k_L, 0.3\,\hbar k_L]$ and $\mathcal{E} = [0,0.1]$, is defined as follows:
\begin{widetext}
\begin{align}\label{Eq:Cost}
    Cost(|\psi(t= t_f)\rangle) =& \bigg\langle\Big|0.5-|\langle +2 \hbar k_L+ p|\psi(t= t_f)\rangle_{\epsilon_{pol}, p}|^2\Big|+\Big|0.5-|\langle -2 \hbar k_L+ p|\psi(t= t_f)\rangle_{\epsilon_{pol}, p}|^2\Big|\nonumber \\
    &+\Big||\langle +2 \hbar k_L+ p|\psi(t= t_f)\rangle_{\epsilon_{pol}, p}|^2-|\langle -2 \hbar k_L+ p|\psi(t= t_f)\rangle_{\epsilon_{pol}, p}|^2\Big|\bigg\rangle_{\mathcal{E},P},
\end{align}
\end{widetext}
where the cost is averaged over all possible pairs $(p,\epsilon_{pol})$ of the selected uniform distributions in $P$ and $\mathcal{E}$. The OCT BS efficiency is defined by $1-Cost$. The first two terms consider the distance from $50 \%$ population of both $|p+2\hbar k_L\rangle$ and $|p-2\hbar k_L\rangle$ and the last term considers the asymmetry in populations between them.

For the case of $\epsilon_{pol} = 0$, the cost function is the same, but we have only one value in $\mathcal{E}$. The results of the optimization are $\left(\Omega_R, \tau, t_0\right) = (2.079\omega_{rec}, 0.534\omega_{rec}^{-1}, 2.463\omega_{rec}^{-1})$, and the time-dependent detuning can be seen in Fig.~\ref{Fig:TDetuning}. 

Contrary to the case where only polarization errors were considered, we noticed that if we optimize over $\left(\Omega_R,\tau, t_0\right)$, apart from the time-dependent detuning, we obtain a much better result than if we decide to start with the same pulse parameters as in Sec.~\ref{sec_pol_err_mitigation}.

For the case where we have both polarization errors and an initial momentum $p\neq 0$, the results of the optimization are $\left(\Omega_R, \tau, t_0\right) = (1.646\omega_{rec}, 0.788\omega_{rec}^{-1}, 4.770\omega_{rec}^{-1})$, and the time-dependent detuning can be seen in Fig.~\ref{Fig:TDetuning}. 

For the case where we have an incoming wave packet with momentum width of $\sigma_p =0.05\,\hbar k_L$ and polarization errors, the results of the optimization are $\left(\Omega_R,\, \tau,\, t_0\right) = (1.264\,\omega_{rec}, \,0.915\,\omega_{rec}^{-1}, \,4.065\,\omega_{rec}^{-1})$, and the time-dependent detuning can be seen in Fig.~\ref{Fig:TDetuning}. 

 We have written the axis in units of $\omega_{rec}$. To get an idea of the duration of these plots, we will give an example for the pulse shown in Fig.~\ref{Fig:TDetuning}. For the case of $^{87}Rb$ with a wavelength of $780.1\times10^{-9}\,m$, the duration is about $332\, \mu s$.

\end{appendix}


\begin{thebibliography}{99}
\bibitem{footnote_on_omega} For generic time-dependent laser frequencies, $\Delta\omega(t) t =[4\omega_{rec}
+\Delta(t)]t\equiv\phi(t_i)+\int_{t_i}^t[4\omega_{rec}
+\Delta_{lab}(t)] \,dt$ denotes the physically accumulated relative laser phase difference. Hence, $\Delta(t)$ discussed in the main text of the paper stands for a fictitious detuning parameter. However, it has a one-to-one correspondence to the laboratory detuning $\Delta_{lab}(t)=\frac{d\Delta(t)}{dt}t+\Delta(t)$ with a properly adjusted initial relative laser phase $\phi(t_i)=\int_0^{t_i}[4\omega_{rec}
+\Delta_{lab}(t)] \,dt$. Every time-dependent detuning protocol $\Delta(t)$ optimized in the paper corresponds to setting a physical detuning $\Delta_{lab}(t)$ in the laboratory.
\end{thebibliography}
\end{document}